\documentstyle[emulate_apj]{article}
\slugcomment{Submitted to {\it The Astrophysical Journal}}
\lefthead{S. Andreon \& S. Ettori}
\righthead{Is the Butcher-Oemler effect a function of z?}
\begin{document}

\title{Is the Butcher-Oemler effect a function of the cluster redshift?}
\author{S. Andreon}
\affil{Osservatorio Astronomico di
Capodimonte, via Moiariello 16, 80131 Napoli, Italy\\
E-mail: andreon@na.astro.it}

\and

\author{S. Ettori}
\affil{Institute of Astronomy, Madingley Road, Cambridge 
CB3 0HA, UK\\ E-mail: settori@ast.cam.ac.uk}

\begin{abstract} Using PSPC {\it Rosat} data, we measure x--ray
surface brightness profiles, size and luminosity of the
Butcher--Oemler (BO) sample of clusters of galaxies. 
The cluster x-ray size, as 
measured by the Petrosian $r_{\eta=2}$ radius, does not
change with redshift and is independent from x-ray luminosity.
On the other hand, the x--ray luminosity increases with
redshift. Considering that fair samples show no-evolution, or negative
luminosity evolution, we conclude that the BO sample is not formed from the same
class of objects observed at different look--back times.
This is in conflict with the usual interpretation of the Butcher--Oemler 
as an evolutionary (or redshift--dependent) effect, based on the 
assumption that we are comparing
the same class of objects at different redshifts.
Other trends present in the BO sample reflect selection criteria
rather than differences in 
look-back time, as independently confirmed by the fact that 
trends loose strength when we enlarge the sample with x--ray selected
sample of clusters. The variety of optical sizes and
shapes of the clusters in the Butcher--Oemler sample,
and the Malmquist-like bias, are the reasons for these
selection effects that mimic the trends usually interpreted as
changes due to evolution.  \end{abstract}


\section{Introduction}

Galaxies in distant ($z\sim0.4$) clusters
differ from those in the nearby systems (e.g.
Butcher \& Oemler 1984; Dressler \& Gunn 1992). There is
a blueing of galaxy color with redshift,
also known as Butcher-Oemler effect (hereafter BO effect,
Butcher \& Oemler 1977, 1984). 
At $z\sim0.4$ there is a population of almost normal late--type galaxies
that by the present epoch has disappeared, faded or has been disrupted 
(Dressler et al. 1997). 
Distant clusters contain galaxies with disturbed morphologies
and peculiar spectra. The occurrence of these peculiarities
varies from cluster to cluster and, on average, increases with
redshift.  In general, the change of galaxy properties is
explained as the effect of some kind of evolution. 

Oemler, Dressler \& Butcher (1997) proposed a physical reason to
explain why most of the clusters showing a BO effect are at
high redshift and almost none at the present epoch: clusters at
$z\sim0.4$ are much more exceptional objects than present--day
clusters and they are observed in the act of growing by merger
of smaller clumps, in agreement with a hierarchical growth of
structures as described, for example, by Kauffmann (1995). 
Furthermore, this scenario
permits the existence of dynamically young
local clusters, such as the spiral rich Abell 1367 and
2151 clusters, and evolved clusters at high redshift, such as
Cl 0024+16.  In the Oemler et al. (1997) interpretation of the
BO effect, clusters at higher redshift are dynamically
younger, on average, than the nearby ones, because we are
looking at the epoch of an enhanced cluster formation. 

Allington-Smith et al. (1993) showed that galaxies in groups do
not evolve (except passively) and suggest that the BO
effect should be interpreted as an evidence of the important
role played by the cluster environment: evolution is strong in
clusters and negligible in groups. However this idea have been
questioned: Rakos \& Schombert (1995) show that it is difficult 
to fade the majority of the cluster population at
$z\sim0.7$ to make their blue population as scarce as in
present--day clusters. Andreon, Davoust \& Heim (1997) and Ellis
et al. (1997) show that cluster ellipticals and lenticulars are
old galaxies already at $z\sim0.4$, and that the majority of them
cannot be the end product of the blue galaxy population.
Recently, this result has been extended with clusters up to $z\sim0.9$
(Stanford, Eisenhardt \& Dickinson 1998).

Andreon et al. (1997) have made a detailed comparison of the
properties of galaxies in the nearby Coma cluster and the
distant cluster, Cl 0939+47. They found that the spiral population
of these two clusters appears too different in spatial, color
and surface brightness distributions to be the same galaxy
population observed at two different epochs. The Coma cluster is
therefore unlikely to be representative of the end of the evolution 
path of Cl 0939+47.

\medskip 
In order to quantify the effect of evolution on the properties 
of a given class of objects, it is required that the observed class
is the same at different times. In the case of the evolution of  
galaxies in clusters, it is necessary that the high redshift clusters
studied are the ancestors of the investigated
present--day clusters.

The goal of this paper is to test whether the distant and nearby
clusters of the BO sample are really the same population seen at
different epochs or two different populations; in other words, we 
want to check if we are comparing unripe apples to ripe oranges in
understanding how fruit ripens!. We achieve this goal by means of the
X--ray properties of the clusters of galaxies, whose evolution is known.

The paper is organized as follows: in the next section, we
present our sample of clusters. In Section~3, we discuss the x--ray
analysis of their images. The observed trends and their
relevance on the BO effect are presented in Section~4 and 5,
respectively. In Section~6, we summarize our main results. 
 
In the following analysis,
we adopt $H_0=50$ km s$^{-1}$ Mpc$^{-1}$ and $q_0=0.5$. 
The conversion to
the physical dimension is done through the equation (Sandage 1988):
\begin{equation}
{\rm r (kpc)} =  2.91 \ 10^3 \times \theta
\frac{ z + 1 - \sqrt{z +1}  }{H_0 \ (1+z)^2},
\end{equation}
where $\theta$ is the angular radius in arcsec.

\section{Our sample: the Butcher \& Oemler (1985) sample}

The clusters most frequently compared for measuring the BO
effect are listed in Butcher \& Oemler (1985). This list is
the master list for many studies (e.g.
Dressler \& Gunn 1992; Oemler, Dressler \& Butcher 1997; half of the 
sample by Smail et al. 1997; Dressler et al. 1998 is drawn from the BO
sample, etc.).

Our sample is made from the BO sample plus
Cl 0939+47, a cluster at $z\sim0.4$ which is frequently studied in the
context of the BO effect. This sample, which consist of 33(+1)
clusters, is not complete in any sense (in cluster richness, in
$z$, etc.). 
Figure 1 reproduces Figure 3 in Butcher \& Oemler (1984), with
the addition of Cl 0939+47. The BO effect is evident from the
increase of the fraction of blue galaxies with the redshift. 

We have x--ray data for 30 of the 34 clusters. Position Sensitive
Proportional Counter (PSPC) images are available for 25 of
these. We call this subsample ``HQ" (high quality) sample.
Five more clusters, as well as many of the clusters observed by {\it
Rosat}, have been observed with previous x--ray missions. We use these old
data when necessary. 

Table 1 presents the whole sample in order of
increasing redshift. A double tick and a single tick in the last column 
indicate that the cluster belongs to the HQ sample and we
collect x--ray flux from literature, respectively. 
Columns 1, 2, 3, 4, 5 and 6 list respectively the cluster name, the
cluster redshift,
the radius $r_{30}$ that contain 30 \% of the whole cluster population,
the number of galaxies $N_{30}$ inside such a radius, and the cluster blue
fraction $f_b$ (from Butcher \& Oemler (1985)). Columns 7 and 8 list the
Galactic HI column (from Stark et al. 1992) toward the cluster direction
and the cluster richness [from the Abell, Corwin \& Olowin (1989,
hereafter ACO) catalog].  We update the richness classification of the
Cl 0939+47 (Abell 851) and Abell 370 clusters, and we attribute a richness
to Cl 0024+16 by adopting the more accurate values listed in Oemler et al.
(1997). Cl 0016+16 is twice as rich as Coma (Koo 1981).  The ACO richness,
according to its definition, is the background corrected number of
galaxies within 3 Mpc from the cluster center having luminosity in the
range $M_3$ and $M_3+2$, where $M_3$ is the magnitude of the third
brightest cluster galaxy.  

\section{Data analysis}

\subsection{Data reduction}

PSPC images in the hard band 0.5--2 keV with 15 arcsec pixel size have
been extracted from the public archive at the Max--Planck--Institut
f\"{u}r extraterrestrische Physik (MPE) or at the Goddard Space Flight
Center (according to their availability). Table 2 lists x--ray related
quantities. We correct our images for exposure variation and telescope
vignetting using the distributed exposure maps. All pixels contaminated by
other objects or occulted by ribs have been flagged and excluded from the
following analysis. 

In order to measure the x--ray radial profiles, brightnesses are computed
in elliptical annuli of semi-major axis increasing in geometrical
progression of base $\sqrt{2}$ in order to take the S/N approximatively
constant along the radius. 

Ellipticity, position angle (PA hereafter) and center for the cluster
emission have been derived paying attention to the observational data
available for distant clusters. For example, we keep fixed the center,
even if isophotes center moves, since in distant clusters we seldom
have data of good enough quality to measure the displacement of the
center of the various isophotes. 
When data are not good enough to estimate the ellipticity or the position
angle, we adopt circular apertures. 

The details of the data reduction are as follows:

-- The equivalent radius of each ellipse is that of a circle of the
same area, i.e. $r=\sqrt{ab}$, where $a,b$ are the major and minor axis of
the ellipse. 

-- The axis lengths of an elliptical annulus of finite width are at half
way from the internal and external edges. 

-- The brightness in an annulus is computed as the ratio of the intensity
measured in unflagged pixels and their total area. 

-- We assume that flagged pixels have the same brightness as unflagged
ones in the same annulus. The flux inside an ellipse is the sum, over the
internal annulus, of the product of the brightness computed in each
annulus and the total area of the annulus (calculated including the
flagged pixels). 

Before we proceed further in the data analysis we need to verify two
assumptions: the computed profiles are independent of the exact choice of
the flagged pixels and of the ellipticity and PA chosen for the
integration. We apply two different flag schemes to the same image of
Abell 2218: (i) we flag only superposed objects and ribs, (ii) we flags
every small fluctuation, including very faint ones at the level of the
noise.  The resulting profiles are indistinguishable. Further confirmation
of the independence of the exact choice of the pixels flagged came from
the comparison of independent analysis of the same images (see next
section). 

In order to test the sensitivity of our profile to the chosen axis ratio,
we compute the profiles of Abell 2218 within elliptical annuli of axis
ratio 0.83 and 1. The two profiles are, again, indistinguishable. The
profiles of some other clusters, taken from the literature, measured
through ellipses of different shapes agree within the errors. Our
elliptical profiles do not depend upon the adopted axis ratio, due to our
selection of the axis length, and the fact that the PA and ellipticity
of the x--ray isophotes are not subject to large variations. 

Furthermore,
we adjust the background level of the Abell 1656 cluster, the
emission of which almost fills the PSPC field of view, in such
a way that our profile at large radii matches those derived in literature
from the {\it Rosat} All Sky Survey images (Briel, Henry \& Boeringer
1992). 
Abell 400 exhibits a central point source, a dumbbell galaxy also known as
radio source 3C75, with a slight offset with respect to the center of
the cluster emission (cf. Beers et al. 1992). Pixels affected by this
source have been flagged.  Some other clusters in our list have some
peculiar features in their x--ray profile: Abell 1689, 2199 and 2634
present a strong cooling flow (Allen \& Fabian 1998, White et al. 1997 and
Schindler \& Prieto 1996, respectively). In the following figures, we
represent with squares symbols these cooling flows clusters. 

\subsection{Comparisons with previous {\it Rosat} data}

A comparison between our profiles and those from literature is quite
difficult.  For most of the clusters, the data points of the surface
brightness profile are not available. Generally, the
best--fit parameters for a $\beta$-model (Cavaliere \& Fusco-Femiano
1976) are the only quantities 
quoted. We will use these for our comparisons, even if some information
is still missing, such as (i) the adopted center, ellipticity and position
angle (for elliptical profiles), (ii) the value of the central brightness,
(iii) the radius up to which the data are fitted, (iv) how well the model
describes the profile, in terms of the location of the deviations from the
best fit. Finally some literature values are wrong through mistake or
typographical errors. 
 
Figure 2 shows good agreement between the best fits as obtained from
literature, once the necessary (if any) corrections were introduced, and
our profiles. Here we note that where we see a local mismatch
between the data and the best fit, the same deviations are often
observed in the published surface brightness profile.

\subsubsection{Remarks on individual clusters}

Abell 262:  David, Jones, \& Forman (1996) found that the x--ray isophotes
of this cluster twist and ellipticity changes with radius. They present
a detailed analysis of the x--ray profile computed through elliptical
apertures whose PA and ellipticity are fitted to the cluster isophotes.
However, they list only counts integrated within ellipses of unspecified
PA and ellipticity, both of which are probably changing with radius. 
In our comparison, to compute the surface brightness profile, we 
calculate the gradient of the integrated flux in each ellipse and make the
approximation that it was computed within ellipses with the same centre
and axis ratio of 0.8.  
Our profile matches exactly the one from literature at $\log(r)>2.4$.
The agreement is satisfactory at smaller radii given the approximation
involved in the comparison.

Abell 401: Our points match well those of Buote \& Canizares (1996)
$\beta$ model. Another observation performed 6 months later is in good
agreement with the plotted profile, confirming also the temporal stability
of the {\it Rosat} PSPC. 

Abell 1656: Our points lie on the best-fit $\beta$ models in Buote \&
Canizares (1996) and Briel, Henry \& Boeringer (1992).

Abell 2199: This has been observed twice with a temporal gap of 3 years.
As for A401, the two profiles are in good agreement between them and with 
Buote \& Canizares (1996) $\beta$
model. 

Abell 2256: This is a quite studied merging cluster (Markevitch \&
Vikhlinin 1997, Buote \& Canizares 1996; Briel, Henry \& Boeringer 1992). 
Our center is not located on the peak emission but on the barycenter of
the x--ray emission. This explains the rising profile at small radii and
the slight differences with the fit functions from literature. 

Abell 2634: This cluster presents a strong cooling flow, and is not
described at all by a $\beta$ model for $\log(r)<2.4$. At larger radii,
where the profile matches the $\beta$ model, our data are consistent
with the best-fit $\beta$ model in Schindler \& Prieto (1996).

Cl0939+4713: Our profile agrees well with the Schindler \& Wambganss
(1996) $\beta$ profile.

\subsection{Characterization of the cluster profiles}

Usually x--ray profiles are characterized through some parameters,
resulting from a fit to the data of an appropriate function, generally a
$\beta$ model. The use of this model presents some problems. Firstly, this
method is parametric and introduces the width of the bin in the
extracted profile as no-physical scale. Secondly, the $\beta$ parameter,
often referred to as the `slope', is not properly
the slope of the profile at large radii, as one can verify calculating the
radial gradient of the $\beta$ model or, more simply, plotting two
profiles with the same $\beta$, but different core radii. Thirdly, the
best-fit parameters are generally a function of the amplitude of the
errors. 

For these reasons, we prefer to characterize cluster x--ray profiles
through a no-parametric way, computing Petrosian (1976) quantities. A
detailed and recent presentation of Petrosian quantities can be found in
Sandage \& Perelmuter (1990). Briefly, the Petrosian radius $r_{\eta}$
is defined as the radius where the surface brightness {\it at} at that
radius is $\eta$ times fainter than the surface brightness {\it inside}
that radius. Figure 3 shows the surface brightness ($SB(r)$)and the
$\eta(r)$ profiles for a King profile, where $\eta(r)= SB(<r)/SB(r)$. 

Choosing a value for $\eta$, of say 2, the corresponding radius
$r_{\eta=2}$  is completely determined (in our example $\log(r)\sim2.8$).
The Petrosian
radius, as a ratio between two surface brightnesses, does not depend
from quantities that usually affect surface brightnesses, such as Galactic
absorption, cosmological dimming, K-correction and even luminosity
evolution if it is the same at all radii. It could be shown (Petrosian
1976), that the Petrosian radius is a metric radius, i.e. its angular
dimension is given by the formula relating the physical dimension of a
rigid rod and its angular dimension. For objects with profiles of the same
shape, the luminosity within a fixed Petrosian radius gives a fixed
fraction of the total luminosity, as the effective radius for the de
Vaucouleurs' (1976) law.  
We choose $2.5\log\eta=2$, and we refer to it as ``$\eta=2$".  
For Hubble and $\beta$ model (with $\beta=2/3$) profiles, $\eta=2$ 
correspond to 55 and 38 core radii, respectively. 

\subsection{Luminosities \& errors}

The count rates have been converted to the flux in the 0.5-2 keV band
using a conversion factor of $1.15 \times 10^{-11}$ erg s$^{-1}$ cm$^{-2}$
/ (count s$^{-1}$), almost independent from the gas temperature.
The correction for the Galactic absorption has been calculated applying
the Morrison \& McCammon (1983) model as a foreground absorber to the 
thermal emission from the intracluster
plasma with metallicity fixed to 0.3 (Raymond, \& Smith 1977;  up to date
version 1992 in XSPEC version~10).  Since all clusters are at high
Galactic latitude, this correction is small.  
K--corrections have been computed individually assuming 
thermal cluster spectrum. Temperatures have been taken
from White, Jones \& Forman (1997).
For the clusters Abell 222, 223, 777, 963, 1758, 1904, 
2125, and Cl 0024+16, Cl 0939+47, that are not listed in White et al. 1997,
we adopt a temperature of 4 keV. 
Our K--corrections are compatible with the more accurate values
plotted by Jones et al. (1998) in their Figure 7.
Differences amount to 0.01 in $\log(L_X)$ at most, i.e. are
negligible.

Our estimate of the uncertainties include Poisson errors and a generous
10\% error on the determination of the background level. In Figure 4, and
subsequent plots, we do not plot the errors on the x--ray flux, since they
are smaller than the symbol size, except for two clusters (Abell 777 and
Cl0024, whose fluxes are lower limits). 

\subsection{Comparison with data from previous x--ray missions}

For the clusters of our sample, the x--ray luminosities measured by
previous missions are
listed in various compilations (Salten \& Henry 1983; 
Lea \& Henry 1988; Mushotzky \& Scharf 1997; Sadat, Blanchard,
Guiderdoni \& Silk 1998). Their luminosities are not measured at the 
Petrosian radii, nor in the {\it Rosat} hard band, but are simply aperture
or isophotal fluxes, usually in the band of observation. We convert them
in our system (flux measured in the Petrosian $r_{\eta=2}$ radius in
{\it Rosat } hard band) empirically, by means of
the median difference between the (log of the) luminosities in
common clusters. Our fluxes correlate well with literature ones
transformed in our system, as shown in Figure 4. The large
scatter is due to the heterogeneity of literature data and to the
transformation from one band to another, since the formal error on the
x-ray flux in our system is smaller than the point size. 
The two outliers 
refer to the cluster Abell 400, whose central emission has been masked out
in our flux measure (see Section 3.2.1), but not in the two
estimates from literature.

\section{Results: the trends}

The aim of this section is to show the existence of trends between
quantities related to clusters properties (richness, size, distance,
x--ray flux, etc.), and to understand the role played by selection
effects on these trends.

\subsection{Size}

Table~3 quotes the $r_{\eta=2}$ size of the HQ sample. 
All clusters, spanning a large redshift range, from $z\sim0$ to
$z\sim0.6$, have similar sizes of $\log r \sim 3.10 $ kpc with a scatter
(in $\log r$) of only 0.14 (see Figure 5).  The outliers (at small
$r_{\eta=2}$) appear to be cooling flow clusters.  Cl 0024+16 and Abell
777 have a too noisy profile to compute $r_{\eta=2}$. Our results confirm
those obtained from Henry et al. (1979) and Vikhlinin et al. (1998).
Figure 6 shows that clusters have
similar size, independently on their x--ray luminosity, at least in the
luminosity range sampled ($43.5<\log L_X<45.5$ erg s$^{-1}$).  Furthermore
clusters rich in blue galaxies (solid dots in the figure) are not
preferentially larger, smaller, brighter or fainter than those clusters
poor in blue galaxies. 
 
Figure 7 compares the optical cluster radius, defined as the radius which
encloses 30 \% of the galaxy population, $r_{30}$, with our x--ray
$r_{\eta=2}$ size, for the HQ sample.  The $r_{\eta=2}$ is in
average $\sim3$ times larger than $r_{30}$, with a large scatter. Clusters
rich in blue galaxies (solid dots) do not have systematically larger or
smaller $r_{\eta=2}/r_{30}$ ratios than clusters poor in blue galaxies
(open dots). Even if the two most distant clusters have both a
$r_{\eta=2}/r_{30}$ ratio larger than the average, there is no
convincing statistical evidence for a trend of an increasing
$r_{\eta=2}/r_{30}$ ratio with redshift. 

\subsection{$L_x$ vs $z$}

Figure 8 shows that in the BO sample there is a deficit of distant
clusters with a x--ray luminosity comparable to faint present--day
cluster, and an excess of clusters which are as bright as, or brighter
than, the brightest nearby clusters. This holds for the HQ sample 
as well as for the whole sample.
The x--ray luminosity is correlated with $z$ at the 99.9 \% confidence level,
according to the Spearman's $r_s$ and Kendall's $\tau$ tests.
The x--ray luminosity of 
the four clusters not present in the HQ sample has been converted to our
energy band as described in Section 3.5.
In the whole sample, clusters rich in blue galaxies (solid dots)
are not preferentially the brightest or the faintest ones. The correlation
between x--ray luminosity and redshift is still present if we use $r_{30}$
or a 3 Mpc aperture for all clusters. If we remove the irregular clusters 
identified by Butcher \& Oemler (1984) from the sample the correlation
is still present, but only at 98.5 \% confidence level.

In the x--ray waveband, distant clusters are not brighter in the past
than today, and, if anything, they were fainter in the past, not
brighter (Henry et al. 1992; Collins 1997;  Vikhlinin et al. 1998;
Rosati, Della Ceca, Norman et al. 1998). On the other hand, the x--ray
luminosity of the clusters in the BO sample, that span the same redshift
and luminosity range of the above-mentioned {\it representative}
samples, increases with redshift (or low luminosity clusters are   
missing at large $z$ in the sample). This means that the BO sample is
not representative of a homogeneous class of clusters of galaxies
observed at different look--back times, but it is biased toward an
increasing fraction of bright x--ray clusters as the redshift increases.
We postpone the discussion of the relevance of the trend in the BO
effect to the next section. 

The existence of a strong correlation between x--ray luminosity and
redshift, in the BO sample, makes suspicious any other correlation
involving these two quantities. 

\subsection{Richness vs $z$}

In hierarchical scenarios, clusters at high redshift are more massive, on
average, than nearby clusters, since only richest clusters are already
formed. Instead, the ancestors of present--day clusters were less massive
than today and they had not yet formed at high redshift. Therefore, it is
expected that, at higher redshift, clusters (which have already formed)
are richer than nearby ones. 

Figure 9 shows that the distant clusters in our sample are
also the ones with higher ACO richness. The Spearman $r_s$ and Kendall's
$\tau$ tests reject ($>$ 99.9 \% confidence level) that the
richness is not correlated with $z$. Dressler et al. (1997),
in their study of the morphological segregation in clusters at $z\sim0.4$
(half of these taken from the BO list), noted that the distant clusters
are denser than nearby ones listed in Dressler (1980).  

The increase of the cluster richness with redshift in our (BO)
sample is not due to the evolution of clusters, but just to two
selection effects: both richness and x--ray luminosity (see Figure 10),
and x--ray luminosity and redshift (see Figure 8) are correlated. 
The latter correlation is certainly a bias, and this induces a
correlation between redshift and richness.
Therefore the trend between richness and redshift is not a property of
the clusters but a result of the (poorly known) selection criteria adopted
for assembling the sample. 

The (apparent) evolution of the cluster richness is easy to
understand from an observational point of view. In the optical, clusters
are usually detected as galaxy overdensity over the field. As
the redshift increases, the clusters have to be richer and richer to be
detected, and distant poor clusters are likely to be missing in all
optically selected catalogs. The ACO catalog, on which the
Butcher--Oemler sample is largely based (note also that Cl 0939+47 is
listed in the ACO catalog as Abell 851), is complete up to $z\sim0.1$
(Scaramella et al. 1991). At larger redshifts only the richest clusters
are present, whereas at small redshift the number of very rich clusters is
small because of the small local volume. 
 
The right panel of Figure 9 shows that the central richness, $N_{30}$, of
clusters with PSPC data does not increase with redshift, contrary to
that expected from its correlation with x--ray luminosity (Figure 10)
and from the increase of the x--ray luminosity with $z$ (Figure 8).
However, it is quite dangerous to do predictions by propagating
correlations between quantities, especially in a biased sample such as
our, since too many properties are changing at the same time as the
redshift varies. 

$N_{30}$ and $R$ do not show any statistically significant correlation
(the Spearman test indicate a correlation at 60 \% confidence level). Poor
clusters, in the ACO sense, do not have too many galaxies within $R_{30}$,
whereas rich clusters can be very rich, as well as very poor, in the
center. This means that clusters have a variety of galaxy density profiles
for a given $N_{30}$ or $R$, since for the same total number of bright
galaxies, $R$, they can have quite different central number of galaxies,
$N_{30}$ (and vice versa). Alternatively, large observational errors 
affect these two quantities. 

\subsection{$L_x$ vs richness}

Figure 10 shows that in the whole sample the cluster x--ray luminosity
increases with galaxy richness, as measured by either Abell, Corwin \&
Olowin (1989) or Butcher \& Oemler (1984). Clusters which are rich in blue
galaxies (solid dots) span the entire range explored
in richness and luminosity.  The correlation between x--ray luminosity
and cluster richness is expected (e.g. Bahcall 1974; Jones \& Forman
1978). However, in our sample, this correlation is probably the result
of two selection effects: as the redshift increases, we sample (i)
brighter (see Figure 8) and (ii) richer (Figure 9, left panel) clusters.  
Our statement can be checked using the data from
Smail et al. (1998), who studied very bright x--ray selected clusters,
independently from their optical richness. Their clusters have $\log
L_x\sim45$, $0<R<3$ and $15<N_{30}<60$.  Adding these data to ours, the
correlation between richness and x--ray luminosity largely disappears,
thus confirming that the found correlation is the result of the
selection criteria instead of a real clusters property (Figure 11). 

\subsection{$L_x$ vs $f_b$}

A correlation between $L_x$ and $f_b$ would explain many
cluster properties. The lack of blue galaxies in the cluster core, the
color distribution of spiral galaxies and many of their properties, such
as velocity and position relative to the cluster center, higher surface
brightness (Andreon 1996) and HI deficiency of infalling spirals (Gavazzi
1987) can be explained if spirals falling in clusters have a starburst
due to the ram pressure in the hot gas (Bothun \& Dressler 1986) that
consumes the galaxy's gas reservoir. During the burst, these galaxies
become bluer and brighter in the mean surface brightness. Just after the
burst, they become as red as ellipticals (Charlot \& Silk 1994),
explaining the presence of red spirals in cluster cores.  Furthermore,
both the existence of galaxies that show spectral signatures consistent
with the presence of intermediate age stellar populations (Couch \&
Sharples 1987, Lavery \& Henry 1988, Dressler \& Gunn 1992), and the
photometric evidence for the blue starburst spirals in Coma (Donas et al.
1995, Andreon 1996), give support to this scenario. 

We do not observe any correlation between x--ray luminosity and the
fraction of blue galaxies for the whole sample and for
the HQ sample. In a sample of 10
clusters at moderate redshift ($z\sim0.25$), which span just a factor
2-3 in x--ray luminosity, a wide spread is found in the fraction of blue
galaxies (Smail et al. 1998), which is
uncorrelated to x--ray luminosity.  Using {\it Einstein Observatory} data,
Lea \& Henry (1988) suggest the possible existence of a correlation
between these two quantities in a sub-sample of the BO list, provided that
deviant points (low luminosity clusters and the most distant cluster) are
discarded. The absence of a correlation between the fraction of blue
galaxies and the cluster x--ray luminosity implies that this link, if
exist, is complex and needs more physical parameters to be explained than
only the spiral fraction and the x--ray luminosity. 

Here we note that these quantities are not averaged on the same cluster
area, nor on regions whose area ratio is fixed: sometimes the optical
radius is 3 time larger than the area over which the spiral fraction has
been computed, and sometimes it is two times smaller (see Figure 7).
For this reasons, we have recalculated the cluster x--ray luminosities
within the radius $r_{30}$ used for computing the cluster spiral fraction,
but still any significant correlation between these two quantities does
not appear. 

\section{Relevance of these trends in the context of the BO effect}

Any sample of local and distant clusters that is not statistically 
complete can be affected by the selection criteria adopted to assemble it.
This happens because clusters have (i) morphological differences 
in the nearby universe (Zwicky 1957) and at $z\sim0.4$ (Oemler et al. 1998) 
and (ii) their galaxy populations are subjected to several segregation effects, 
both in galaxy morphology (Hubble \& Humanson 1922; Dressler 1980; Sanrom\`a \&
Salvador--Sol\'e 1991; Whitmore, Gilmore \& Jones 1993; Andreon 1994,
1996; Dressler et al. 1997; Andreon, Davoust \& Heim 1997), color
(Butcher, \& Oemler 1984; Mellier, Soucail, Fort et al. 1988; Donas et al.
1995; Andreon 1996), and spectral properties (Biviano et al. 1997 and
reference therein).

In particular, any selection done on the basis of the richness is 
contaminated by several factors, such as our ignorance on the physical
evolution of the cluster richness or the role played by local phenomena
as the enhancement in brightness due to starburst activity.
In this sense, selecting clusters according to their x--ray luminosity 
is safer, because the physics of the x--ray emission is well 
known, and easier in detecting (the x--ray emission goes as the square of
the density, instead of the density for optical richness).
Once a sample of clusters is properly defined, the assumption done is 
that the same class of objects are compared at different look--back times.

{\it Our results show that the main cluster sample studied up to now 
in the context of the BO effect is biased:} the x--ray luminosity of these
clusters increases with the redshift, contrary to the recent observational
evidence for representative samples (see Sect.~4.2).
Thus, the nearby and distant clusters in the BO sample are not representative
of a fair sample. This implies that any trend highlighted in the BO sample
could be the product of the selection criteria adopted instead of real
differences with respect to the age of the systems.

Differences in x--ray luminosity reflect, at large extent,
differences in the intracluster gas temperature and gas density
and, consequently, in the cluster mass (Quintana
\& Melnick 1982, Edge \& Steward 1991, White et al. 1997).
Oemler et al. (1997) supposed that they were studying richer and
richer clusters as the redshift increases, and that distant
clusters were growing in a way different from present--day
clusters, i.e. merging smaller clumps at higher rate, as
hierarchical scenarios suggest (Kauffmann 1995).
This conclusion supposes a physical evolution of the clusters
in the BO sample, whereas instead the richness of the
clusters in the BO list increases just because of selection effects.

Another piece of evidence for the presence of a selection bias in
the BO sample comes from the fact that the BO effect is only evident in
optically selected cluster samples. In fact, clusters selected in 
the x--ray band, with almost the same x--ray luminosity and $z\sim0.25$, 
show blue fraction values with a large range, and with a mean similar
to that observed in nearby clusters (Smail et al. 1998). 
This mean value is also smaller than the blue fraction in the BO 
clusters at the same redshift.

To summarize, the BO sample does not contain the same class
of objects at different look--back times, contrary to the requirement to
detect any sign of evolution in a sample.

\medskip
We note that, although this bias affects the BO sample, it could not
lower the significance of the BO effect, if x--ray bright clusters have
the same blue fraction of much fainter clusters. This is a hypothesis
that, at the present time, we cannot test observationally on an unbiased
sample. In the BO biased sample,
the fraction of blue galaxies does not seem to depend on the x--ray
cluster properties. Furthermore, the BO effect is evident even after
removing the faint clusters with $\log(L_X)<44$ and $z<0.1$.
However, we do not know if this reduced sample (or any other
subsample drawn from the BO sample) is representative
for the range of redshift studied, and any conclusion drawn
from it should be regarded with caution. 
In conclusion, we do not believe that selection biases are completely
removed by eliminating offending clusters. 

From the theoretical point of view, Kauffmann (1995) shows that in their
model of cluster formation and evolution the fraction of blue galaxies
does not depend on the cluster mass, at least for rich clusters.
In that case, there is no risk in comparing clusters of
different masses (x--ray luminosities) at different redshift for
studying the BO effect.  We stress, however, that the
evolutionary interpretation of the BO effect still holds only
under the hypothesis that these selection biases do not affect
the sample, hypothesis which must be shown to be true.

Galaxies in groups do not show the BO effect (Allington-Smith et
al. 1993) over the same redshift range.  For this reason, and
under the assumption of the evolutionary interpretation of the
BO effect in clusters, Allington-Smith et al. (1993) claim that
evolution is driven by environment, much more than look--back
time. However, selection criteria of studied groups and clusters
are quite different: groups are {\it not} optically selected,
because Allington--Smith et al. (1993) built their group sample
selecting the galaxies around radio--galaxies of a given radio
flux, which is likely to be uncorrelated with the optical luminosity of
the galaxy hosting the radio--source, or to the group optical
properties. Instead, the  BO cluster sample is biased toward very
rich (and x--ray luminous) clusters at high redshift. We think
that the claim of a differential evolution of galaxies in
clusters compared to those in groups, should be pending on 
a proper determination of the amplitude of the BO effect in a sample
of clusters representative of their redshift. 

\section{Conclusions}

We have analysed {\it Rosat} PSPC images of most of the clusters
studied in relation to the Butcher-Oemler effect. We have
computed surface brightness profiles, as well as x--ray fluxes
within metric diameters adapted to the cluster size (Table 3).  
Our main results are: 

1) The cluster x-ray size, as measured by the Petrosian $r_{\eta=2}$
radius, does not
evolve and it is independent of x-ray luminosity: $\log \
r_{\eta=2}\sim3.10\pm0.14$ kpc. 

2) The x--ray luminosity of clusters listed in the
Butcher--Oemler sample increases with redshift (Figure 8). In the same
redshift range, there is observational evidence, from {\it
representative} samples, that the x--ray luminosity of clusters is 
constant or decreasing i.e. have a trend opposite to those
observed in BO sample. Therefore, nearby and
distant clusters in the BO sample are not representative
of a given class of objects observed at different epochs, and thus
the BO sample does not contain the same class
of objects at different look--back times, contrary to the requirement to
detect any sign of evolution in a sample.
 
Because selection criteria modify the sample composition in a redshift
dependent way, it is quite difficult 
to disentangle a real redshift dependence (evolution) from a fictitious
redshift trend induced by selection criteria. 
Hence, the observed BO effect measured from optical selected samples 
is {\it not necessarily} a general property of clusters of galaxies, but
could be a selection effect.
There is some independent support to this interpretation: it seems that
x--ray selected clusters, all of similar x--ray luminosity and therefore
likely to belong to the same class, do not show the BO effect
(Smail et al. 1998). Similarly, galaxies in radio--selected groups 
show no evolution, beside passive one (Allington-Smith et al. 1993).

The variety of optical shapes and sizes of clusters, together with
the Malmquist-like bias and the incompleteness of the BO
list, are the main sources for the trends present
in the sample. 



X--ray data have been of fundamental importance in revealing 
the existence of a selection bias that mimic the trend usually
interpreted as evidence of evolution. It is not surprising,
therefore, that our conclusions differ from those reached
when x--ray data were not available.

3) The ACO richness of clusters listed in the Butcher--Oemler
sample increases with redshift. We interpret this correlation as
an observational effect: poor clusters are scarcely detected at
high redshift, and, unusually, rich cluster are missing in low
redshift samples. Other cluster quantities ($N_{30}$, $L_X$,
etc.) shows some correlation among them or with redshift. We
explain these as the effect of selection criteria. Adding to our
sample a sample of x--ray selected clusters, the correlations
generally lose strength, suggesting the correctness of our
interpretation. 

4) The usual interpretation of the BO effect, as due to evolution,
holds only assuming
that selection effects have not practical relevance, an hypothesis
which must be tested. The absence of correlation between the
fraction of blue galaxies and the x--ray luminosity of the
clusters may suggest such a possibility.

\acknowledgments 
We are grateful to Sabrina de Grandi for her support and
availability at the beginning of this work and to David White and
Gilles Theureau for their suggestions on the manuscript. Constructive
comments by D. Burstein, G. Bothun and an anonymous referee
improved this paper. 
S.E. acknowledges M. Capaccioli for the financial support 
of the publication charges of this paper.
S.A. dedicates this work to
his grand father, Ismaele Andreon, recently deceased.

\newpage

\begin{deluxetable}{llrrrlrc}
\tablecolumns{8}
\tablewidth{0pc}
\small
\tablecaption{The BO sample}
\tablehead{
\colhead{name} & \colhead{$z$} & \colhead{$r_{30}$} & \colhead{$N_{30}$} 
& \colhead{$f_b$} & \colhead{nh} & \colhead{R} & \colhead{X-ray?}\\
& & \colhead{arcmin} & & & \colhead{[$10^{20}$ atoms cm$^{-2}$]} & & }
\startdata
Virgo & 0.0033 & 120\phd\phn & 21 & 0.04 & \nodata & \nodata & $\surd$ \nl
Abell 262  & 0.0164 & 27\phd\phn & 22 & 0.02 & 5.3 & 0  & $\surd$$\surd$ \nl
Abell 1367 & 0.02 & 25\phd\phn & 20 & 0.4\phn & 2.1 & 2  & $\surd$$\surd$ \nl
Abell 400  & 0.0232 & 17\phd\phn & 30 & 0.05 & 8.7 & 1  & $\surd$$\surd$ \nl
Abell 1656 & 0.0232 & 22\phd\phn & 94 & 0.03 & 0.91 & 2  & $\surd$$\surd$ \nl    
Abell 2199 & 0.0305 & 18\phd\phn & 94 & 0.04 & 0.88 & 2  & $\surd$$\surd$ \nl
Abell 2634 & 0.0322 & 30\phd\phn & 60 & 0.02 & 4.9 & 1  & $\surd$$\surd$ \nl
Abell 2151 & 0.0371 & 14\phd\phn & 29 & 0.14 & 3.4 & 1  & $\surd$$\surd$ \nl      
Abell 2256 & 0.0581 & 11\phd\phn & 116 & 0.03 & 4.2 & 2  & $\surd$$\surd$ \nl     
Abell 1904 & 0.0714 & 9.4 & 68 & 0.02 & 1.8 & 2  & $\surd$$\surd$ \nl
Abell 401  & 0.0748 & 10.7 & 92 & 0.02 & 1.1 & 2  & $\surd$$\surd$ \nl
Abell 2670 & 0.0749 & 4.9 & 51 & 0.04 & 2.7 & 3  & $\surd$$\surd$ \nl
Cl 0004.8-34 & 0.114 & 5.9 & 60 & 0.07 & \nodata & \nodata & \nodata \nl 
Abell 2218 & 0.171 & 5.8 & 114 & 0.11 & 3.3 & 4 & $\surd$$\surd$ \nl
Abell 1689 & 0.1747 & 5.8 & 124 & 0.09 & 1.8 & 4 & $\surd$$\surd$ \nl
Abell 520  & 0.203 & 4.5 & 126 & 0.07 & 7.6 & 3 & $\surd$$\surd$ \nl
Abell 963  & 0.206 & 3.6 & 88 & 0.19 & 1.4 & 0 & $\surd$$\surd$ \nl    
Abell 223  & 0.207 & 3.2 & 67 & 0.10 & 1.9 & 3 & $\surd$$\surd$ \nl   
Abell 222  & 0.211 & 1.6 & 45 & 0.06 & 1.8 & 3 & $\surd$$\surd$ \nl
Abell 1963 & 0.221 & 1.5 & 38 & 0.10 & \nodata & 2  & \nodata \nl 
Abell 1942 & 0.224 & 2.8 & 57 & 0.17 & \nodata & 3 & $\surd$ \nl 
Abell 2397 & 0.224 & 2.0 & 23 & -0.04 & 5.6 & 3 & $\surd$$\surd$ \nl
Abell 777  & 0.226 & 1.4 & 15 & 0.05 & 1.9 & 4 & $\surd$$\surd$ \nl      
Abell 2111 & 0.229 & 4.1 & 155 & 0.16 & 1.9 & 3 & $\surd$$\surd$ \nl
Abell 1961 & 0.232 & 3.4 & 88 & 0.10 & \nodata & 3 & \nodata \nl 
Abell 2645 & 0.246 & 1.4 & 35 & 0.03 & \nodata & 4 & $\surd$ \nl 
Abell 2125 & 0.2472 & 2.3 & 62 & 0.19 & 2.9 & 4 & $\surd$$\surd$ \nl
Abell 1758 & 0.280 & 2.4 & 91 & 0.09 & 1.1 & 3 & $\surd$$\surd$ \nl 
Cl 1446+26 & 0.369 & 0.9 & 42 & 0.36 & \nodata & \nodata & \nodata \nl 
Abell 370 & 0.373 & 2.2 & 107 & 0.21 & \nodata & 2 & $\surd$ \nl 
Cl 0024+16 & 0.39 & 1.1 & 87 & 0.16 & 4.2 & 2 & $\surd$$\surd$ \nl
Cl 0939+47 & 0.407 & 1.0 & \nodata & 0.4\phn & 1.3 & 5 & $\surd$$\surd$ \nl    
3C295 &	0.465 & 1.0 & 45 & 0.22 & \nodata & \nodata & $\surd$ \nl 
Cl 0016+16 & 0.541 & 1.0 & 65 & 0.02 & 4.1 & 4 & $\surd$$\surd$ \nl
\enddata
\end{deluxetable}

\newpage
\clearpage

\begin{deluxetable}{llrclr}
\tablecolumns{6}
\tablewidth{0pc}
\small
\tablecaption{Dataset ID, exposure time, adopted centers, 
axis ratios and PA for the studied clusters}
\tablehead{
\colhead{name} & \colhead{ID} & \colhead{$t_{exp}$\tablenotemark{a}} &
\colhead{center} & \colhead{$b/a$} & \colhead{PA\tablenotemark{b}} \\
& & \colhead{sec} & \colhead{J2000} & & }
\startdata 
Abell 262 & rp800254n00 & 8163 & \phn1~52~47~+36~09~22 & 0.87 & 45 \nl
Abell 1367 & rp800153n00 & 17610 & 11~44~49~+19~41~28 & 1 & 0 \nl
Abell 400 & rp800226n00 & 22203 & \phn2~57~35~+\phn6~00~25 & 0.66 & 30 \nl
Abell 1656 & rp800005n00 & 19819 & 12~59~42~+27~56~34 & 1 & 0 \nl
Abell 2199 & rp800644n00 & 38244 & 16~28~38~+39~32~52 & 0.76 & 135 \nl
 & rp150083n00 & 10063 & 16~28~39~+39~33~07 & 0.76 & 135 \nl
Abell 2634 & rp800014a01 & 9826 & 23~38~29~+27~01~55 & 1 & 0 \nl
Abell 2151 & rp800517n00 & 11341 & 16~04~36~+17~43~21 & 1 & 0 \nl
Abell 2256 & rp100110n00 & 16452 & 17~03~54~+78~38~19 & 1 & 0 \nl
Abell 1904 & rp800257n00 & 3627 & 14~22~16~+48~30~58 & 1 & 0 \nl
Abell 401 & rp800182n00 & 6289 & \phn2~58~59~+13~34~35 & 0.6 & 30 \nl 
 & rp800235n00 & 7009 & \phn2~58~59~+13~34~40 & 0.6 & 30 \nl
Abell 2670 & rp800420n00 & 16554 & 23~54~14~-10~24~53 & 0.74 & 45 \nl
Abell 2218 & rp800097n00 & 39579 & 16~35~52~+66~12~34 & 0.83 & 0 \nl
Abell 1689 & rp800248n00 & 13142 & 13~11~29~-\phn1~20~32 & 1 & 0 \nl
Abell 520 & rp800480n00 & 4565 & \phn4~54~10~+\phn2~55~04 & 1 & 0 \nl
Abell 963 & rp900528n00 & 9989 & 10~17~12~+39~02~40 & 1 & 0 \nl
Abell 223 & rp800048n00 & 6402 & \phn1~37~56~-12~49~08 & 1 & 0 \nl
Abell 222 & rp800048n00 & 6402 & \phn1~37~34~-12~59~23 & 1 & 0 \nl
Abell 2397 & rp800344n00 & 13629 & 21~56~09~+\phn1~23~25 & 1 & 0 \nl
Abell 777 & rp800049n00 & 7464 & \phn9~29~20+78~16~34 & 1 & 0 \nl
Abell 2111 & rp800479n00 & 7028 & 15~39~41~+34~24~52 & 1 & 0 \nl
Abell 2125 & rp800511n00 & 11340 & 15~41~06~+66~16~13 & 0.6 & 135 \nl
Abell 1758 & rp800047n00 & 16142 & 13~32~42~+50~32~54 & 0.7 & 135 \nl
Cl 0024+16 & rp800524n00 & 1069 & \phn0~26~35~+17~09~43 & 1 & 0 \nl 
Cl 0939+47 & rp800102n00 & 13098 & \phn9~43~00~+46~59~31 & 0.74 & 30 \nl
Cl 0016+16 & rp800253n00 & 40325 & \phn0~18~34~+16~26~16 & 1 & 0 \nl
\enddata
\tablenotetext{a}{Exposure times are read in the central region of the
exposure map.}
\tablenotetext{b}{PAs are from North to East anticlockwise}
\tablecomments{The two pointing of
Abell 2199 (Abell 401) have been acquired 3 years (6 month) apart.
Positions listed assumes no pointing errors.}
\end{deluxetable}

\newpage
\clearpage
 
\begin{deluxetable}{llrr}
\tablecolumns{4}
\tablewidth{0pc}
\small
\tablecaption{Results of the analysis}
\tablehead{
\colhead{name} & \colhead{$\log(r_{\eta=2}))$} & 
\colhead{$\log(L(r<r_{\eta=2})$} & \colhead{$\log(L(r<r_{30}))$}\\
& \colhead{[kpc]} & \colhead{[erg s$^{-1}$]} & \colhead{[erg
s$^{-1}$]}}  \startdata
Abell 262 & 2.96 & 43.66 & 43.64 \nl 
Abell 1367 & 3.03 & 43.86 & 43.83 \nl 
Abell 400 & 3.11 & 43.47 & 43.39 \nl
Abell 1656 & 3.10 & 44.60 & 44.56 \nl 
Abell 2199 & 2.83 & 44.39 & 44.41 \nl 
Abell 2634 & 3.15 & 43.78 & 43.79 \nl
Abell 2151 & 3.33 & 43.96 & 43.82 \nl 
Abell 2256 & 3.10 & 44.64 & 44.62 \nl 
Abell 1904 & 3.34 & 43.86 & 43.76 \nl
Abell 401 & 3.26 & 44.80 & 44.77 \nl 
Abell 2670 & 3.03 & 44.17 & 44.09 \nl 
Abell 2218 & 3.05 & 44.73 & 44.75 \nl
Abell 1689 & 2.87 & 45.08 & 45.11 \nl 
Abell 520 & 3.22 & 44.90 & 44.86 \nl 
Abell 963 & 3.31 & 44.82 & 44.68 \nl
Abell 223 & 3.17 & 44.47 & 44.35 \nl 
Abell 222 & 3.29 & 44.49 & 44.13 \nl 
Abell 2397 & 3.12 & 44.60 & 44.45 \nl
Abell 777 & \nodata & $\sim$43.86 & 43.43 \nl 
Abell 2111 & 3.24 & 44.76 & 44.71 \nl 
Abell 2125 & 3.08 & 44.22 & 44.14 \nl
Abell 1758 & 3.10 & 45.00 & 44.94 \nl 
Cl 0024+16 & \nodata & $\sim$44.37 & 44.25 \nl 
Cl 0939+47 & 3.08 & 44.95 & 44.41 \nl
Cl 0016+16 & 3.05 & 45.25 & 45.07 \nl 
\enddata 
\end{deluxetable}

\newpage
\clearpage

\vfill\eject

\begin{figure}
\epsfysize=8cm
\epsfbox[55 200 460 590]{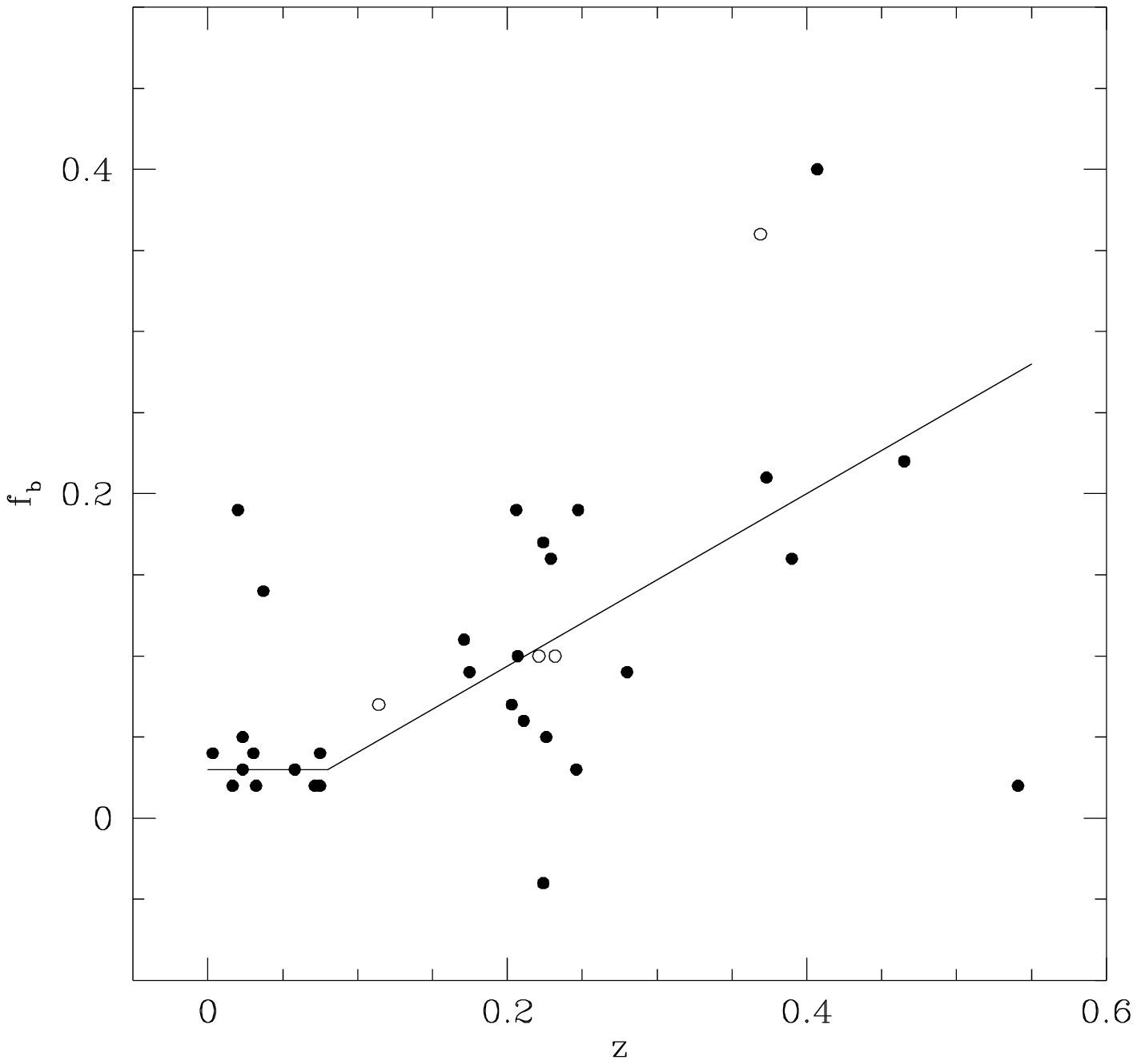}%
\caption[h]{Blue fraction as a function of z for the whole sample (which is the
BO sample with the addition of the cluster Cl 0939+47). Close and open
points mark clusters with and without x--ray data, respectively. The
spline is the Butcher \& Oemler (1984) eye fit to the data.}
\end{figure}

\begin{figure}
\hbox{
\epsfysize=8cm
\epsfbox[60 190 460 590]{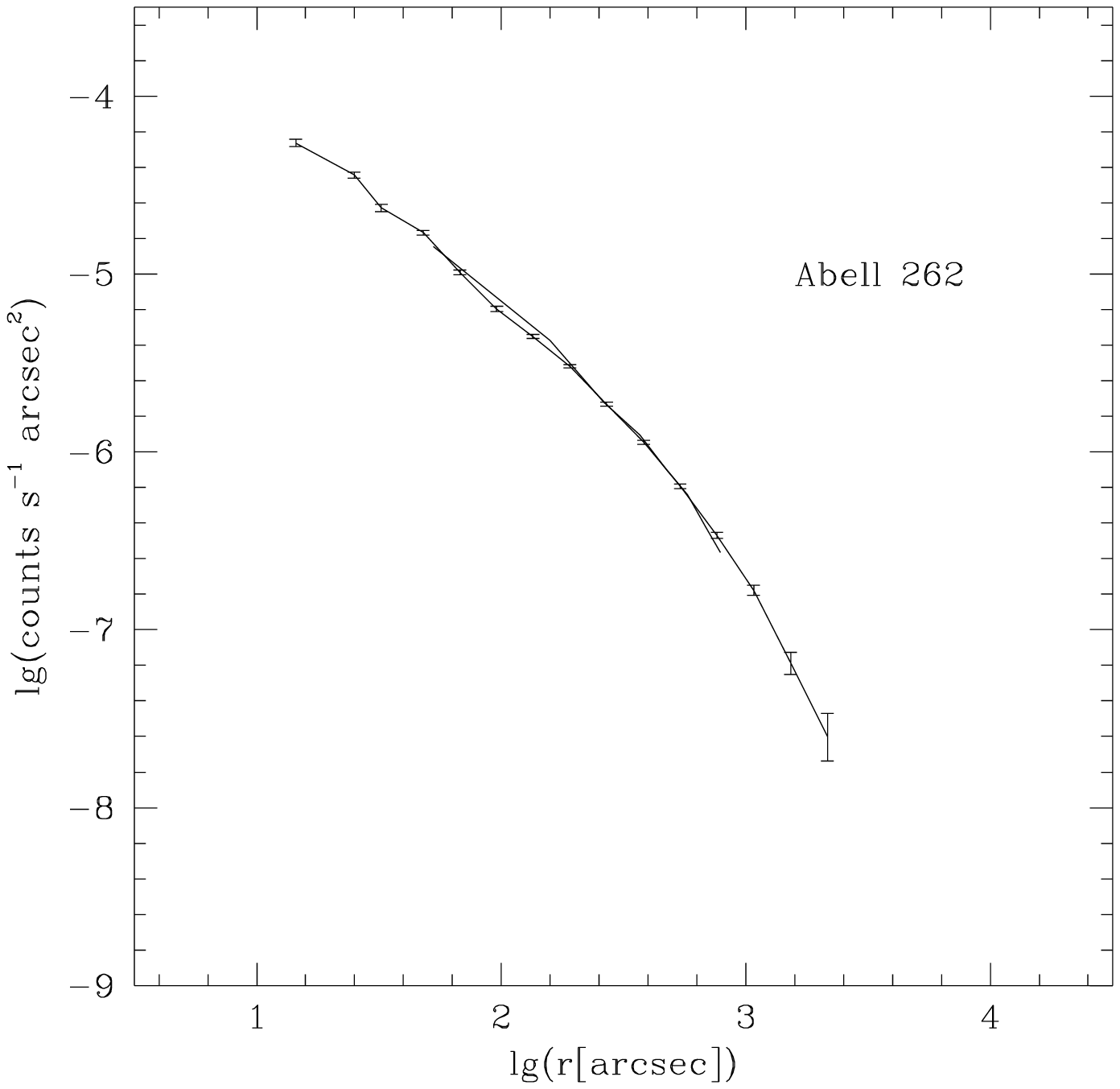}%
\epsfysize=8cm
\epsfbox[60 190 460 590]{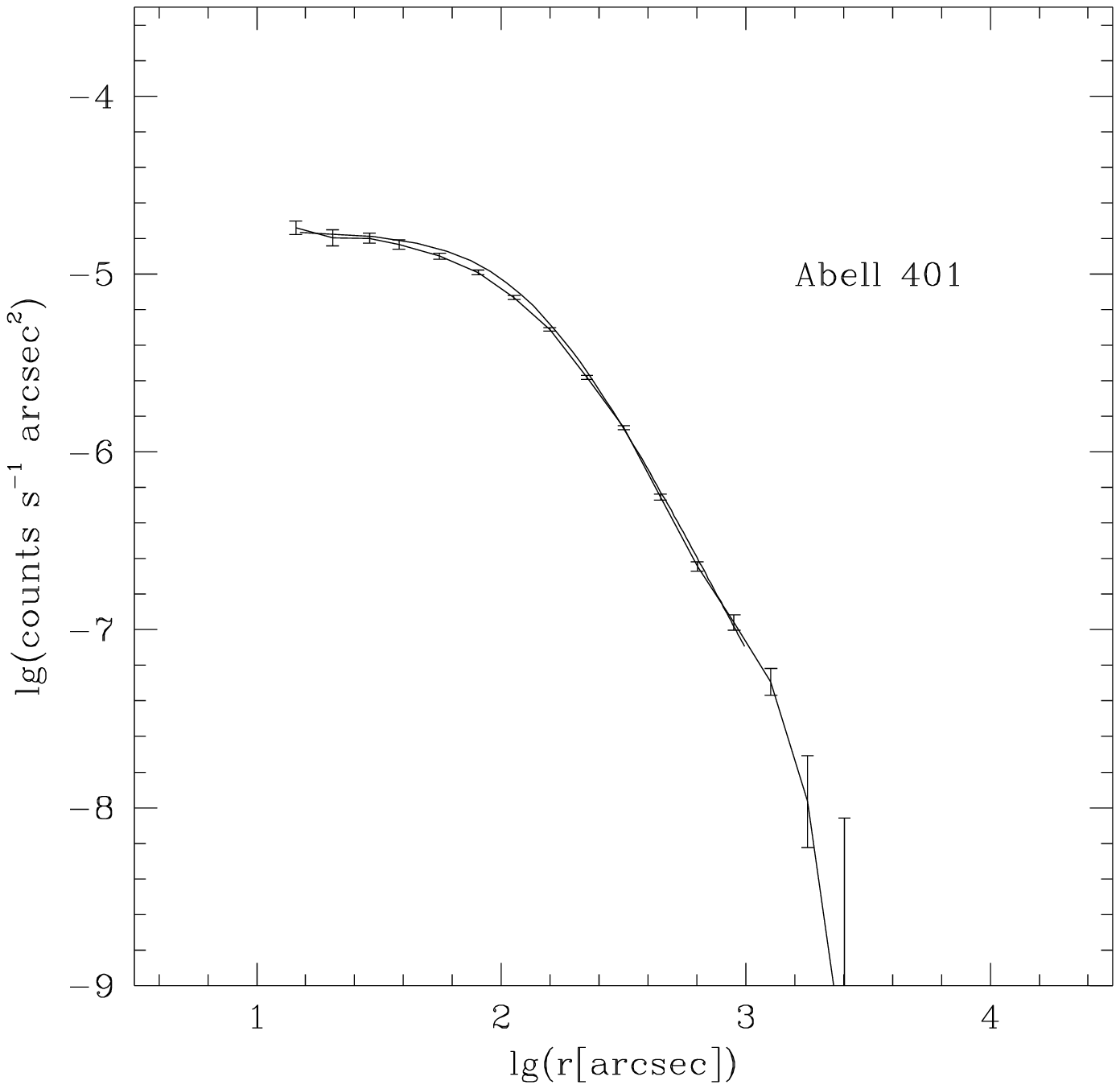}%
\hfill\null}
\hbox{
\epsfysize=8cm
\epsfbox[60 190 460 590]{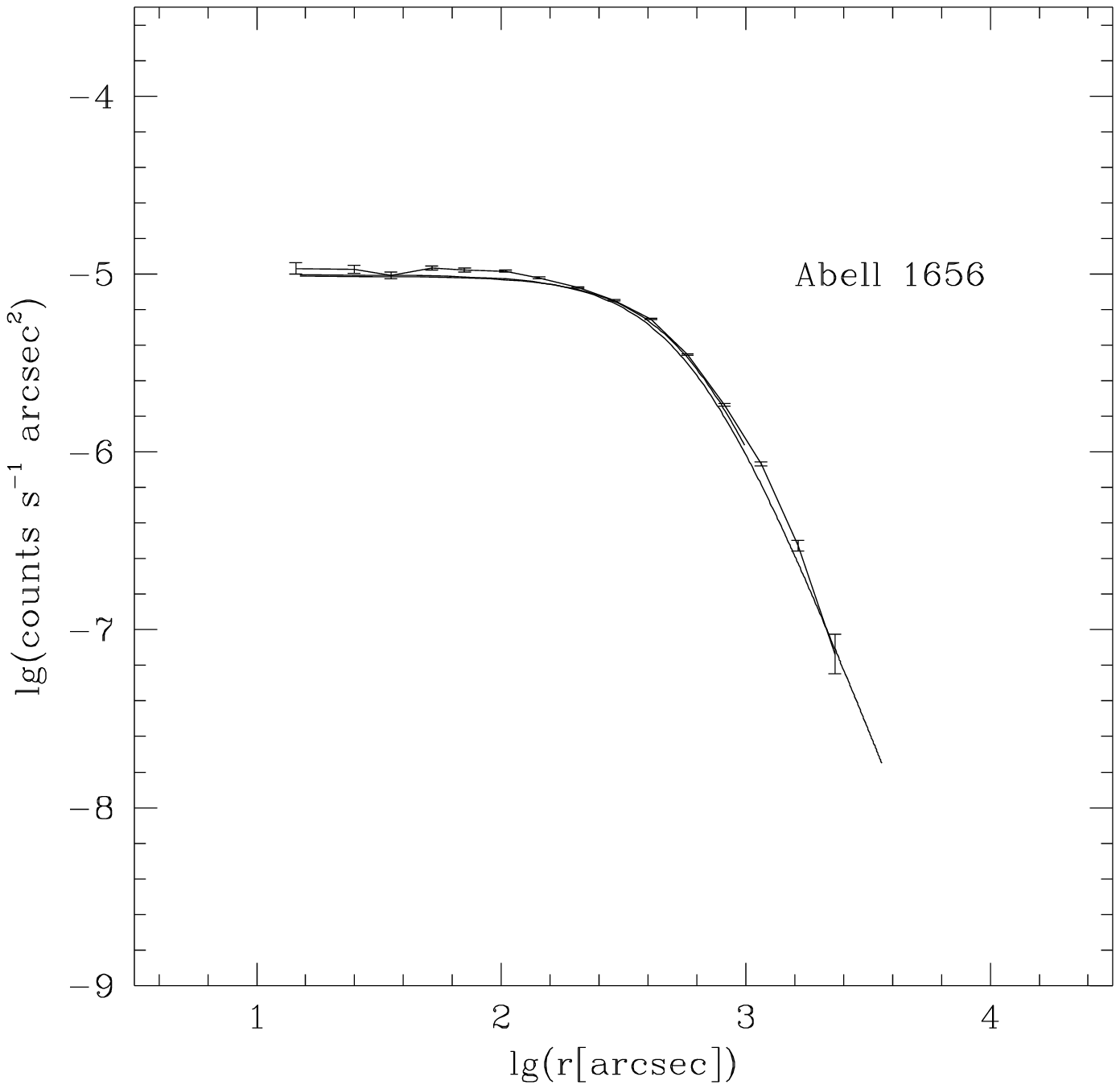}%
\epsfysize=8cm
\epsfbox[60 190 460 590]{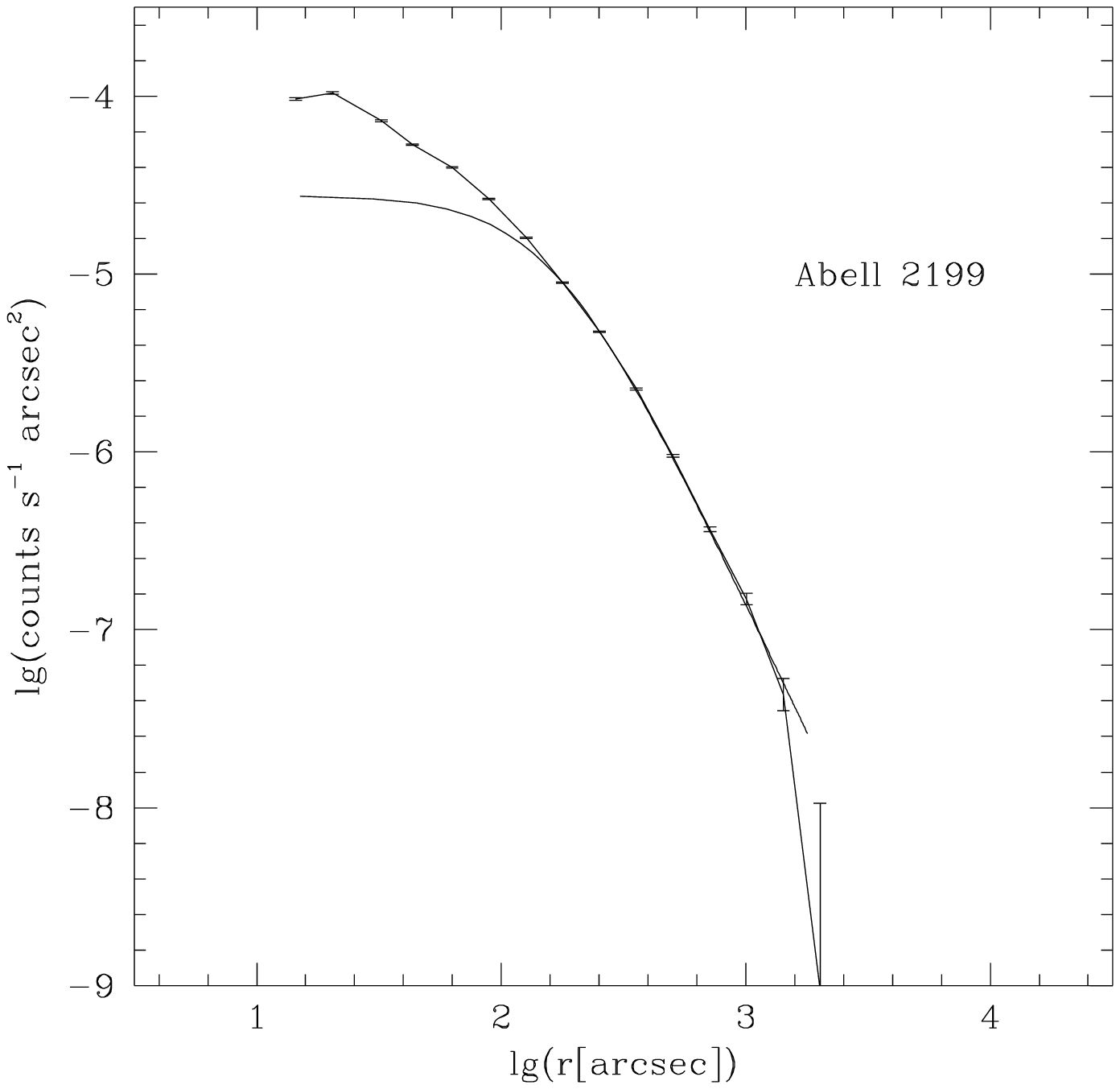}%
\hfill\null}
\hbox{
\epsfysize=8cm
\epsfbox[60 190 460 590]{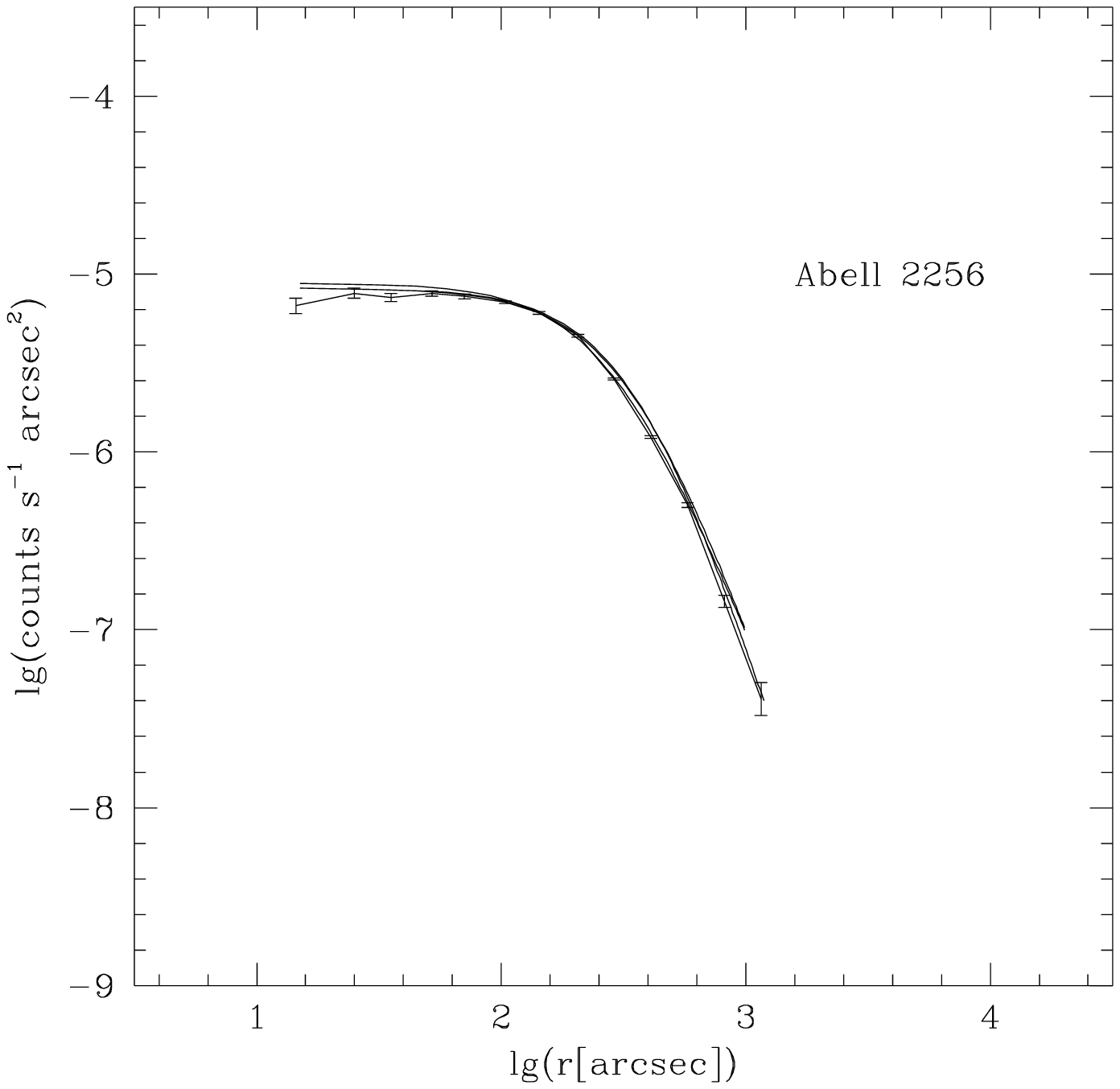}%
\epsfysize=8cm
\epsfbox[60 190 460 590]{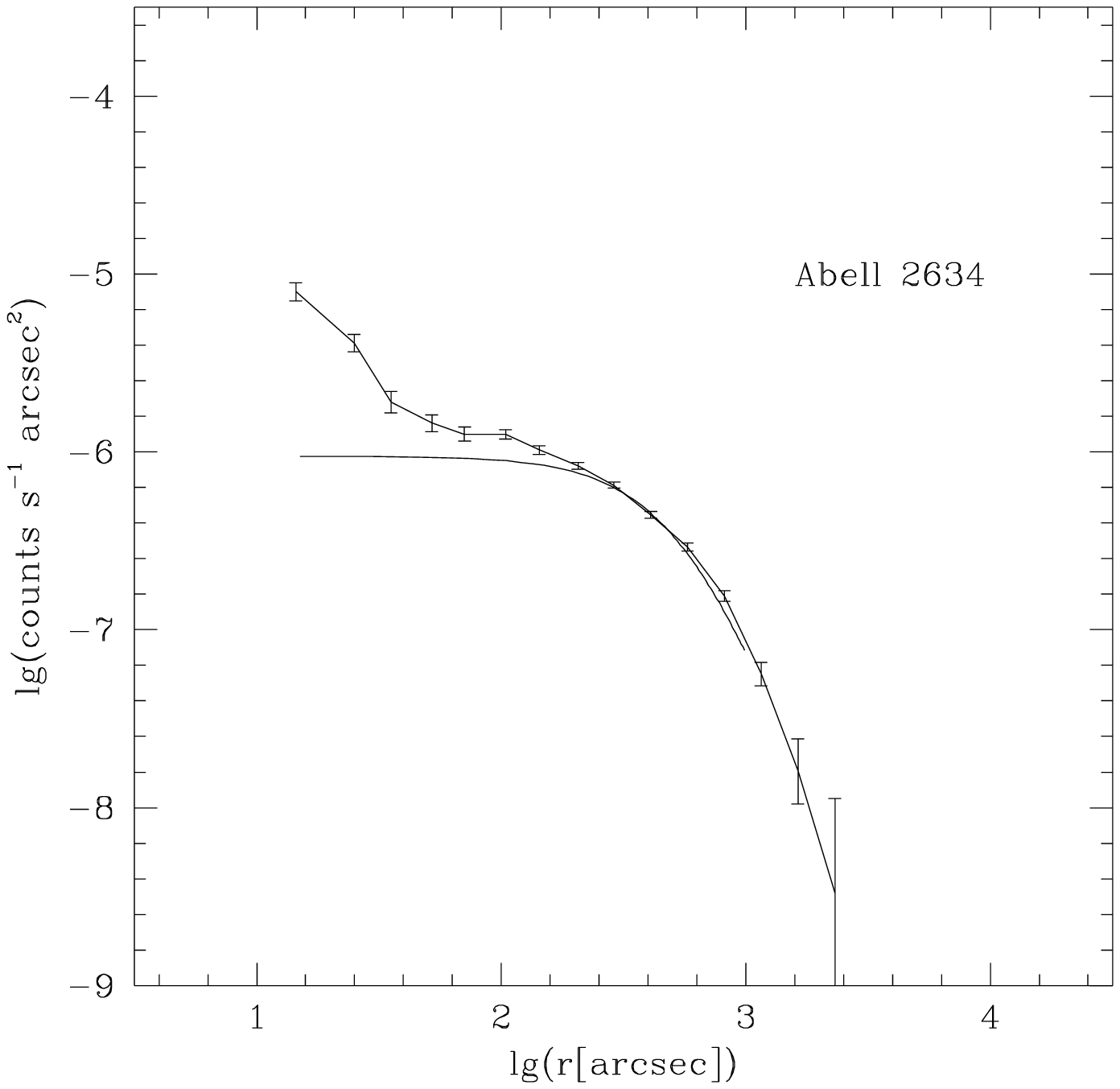}%
\hfill\null}
\end{figure}

\begin{figure}
\hbox{
\epsfysize=8cm
\epsfbox[60 190 460 590]{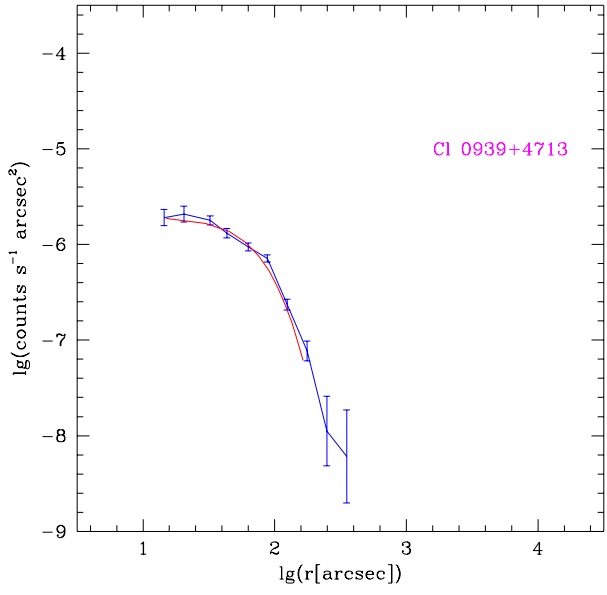}%
\hfill\null}
\caption[h]{Comparison between literature fit and our x--ray profiles}
\end{figure}

\begin{figure}
\epsfysize=8cm
\epsfbox[55 190 490 590]{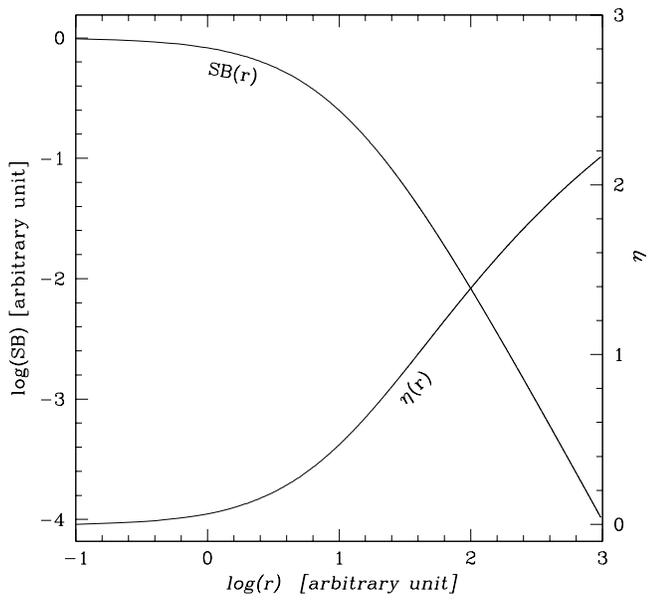}%
\caption[h]{SB (Surface Brightness) and $\eta$ profiles for a King profile
with $\beta=0.5$ and arbitrary core radius and central surface brightness}
\end{figure}

\vfill\eject

\begin{figure}
\epsfysize=8cm
\epsfbox[60 190 470 590]{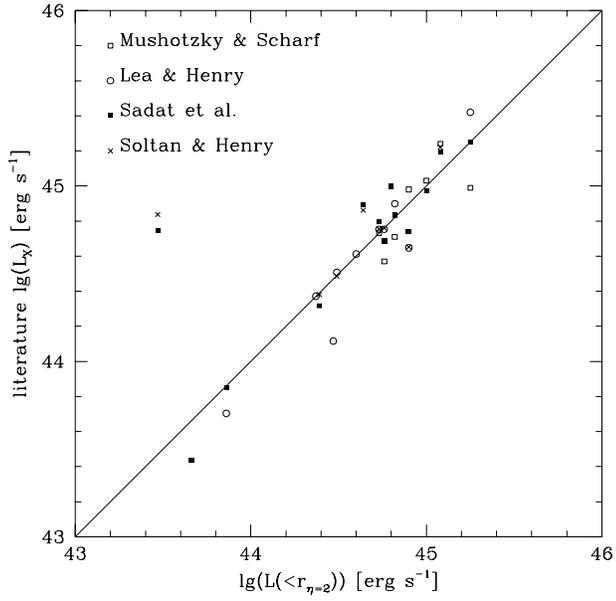}%
\caption{Comparison between our metric fluxes in the {\it Rosat}
hard band and isophotal or aperture fluxes from older satellites
converted in our system.}
\end{figure}

\begin{figure}
\epsfysize=8cm
\epsfbox[40 190 160 690]{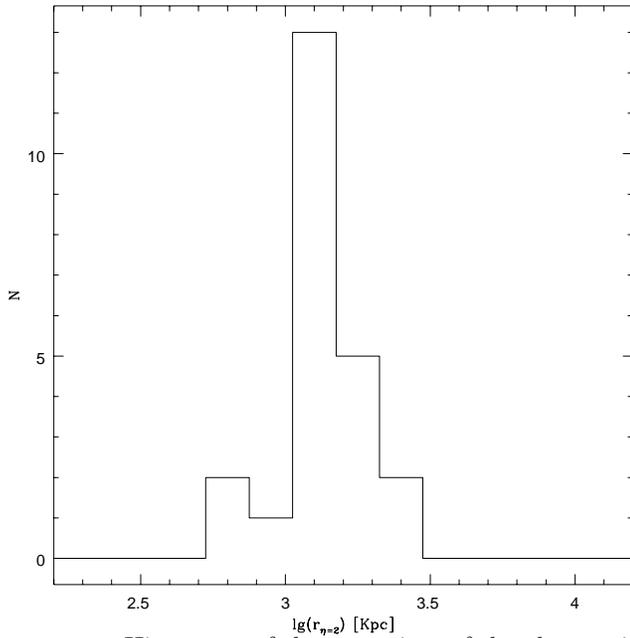}%
\caption[h]{Histogram of the $r_{\eta=2}$ sizes of the  
clusters in the HQ sample.}
\end{figure}

\begin{figure}
\epsfysize=8cm
\epsfbox[40 190 460 590]{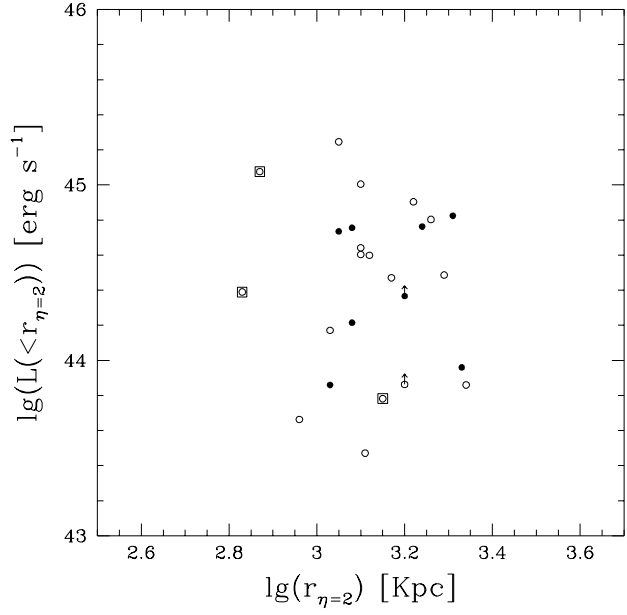}%
\caption{X--ray cluster luminosity as a function of the size for
clusters in the HQ sample.
Solid dots are clusters rich in blue galaxies ($f_b>0.1$), open dots
are clusters poor in blue galaxies. Squares are cooling flow clusters.}
\end{figure}

\begin{figure}
\epsfysize=8cm
\epsfbox[45 200 470 590]{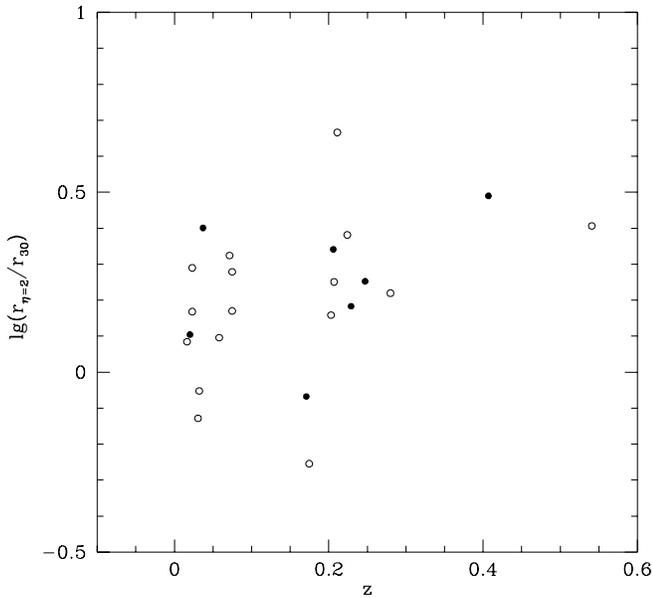}%
\caption{Ratio between the optical radius $r_{30}$ and the x--ray
size $r_{\eta=2}$ as a function of $z$ for
clusters in the HQ sample. Symbols as Figure 6.}
\end{figure}

\begin{figure}
\hbox{
\epsfysize=8cm
\epsfbox[40 190 460 590]{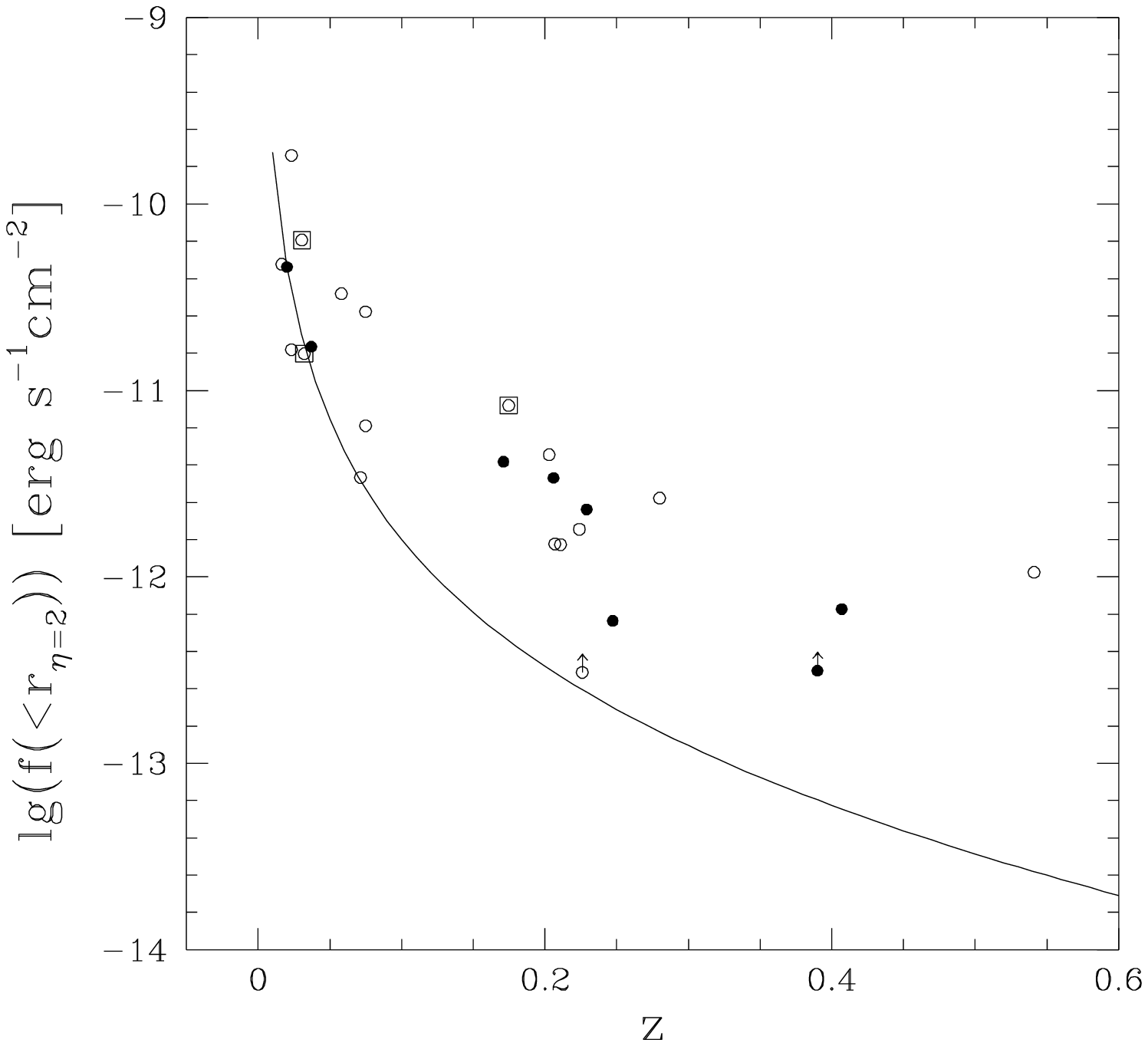}%
\epsfysize=8cm
\epsfbox[40 190 460 590]{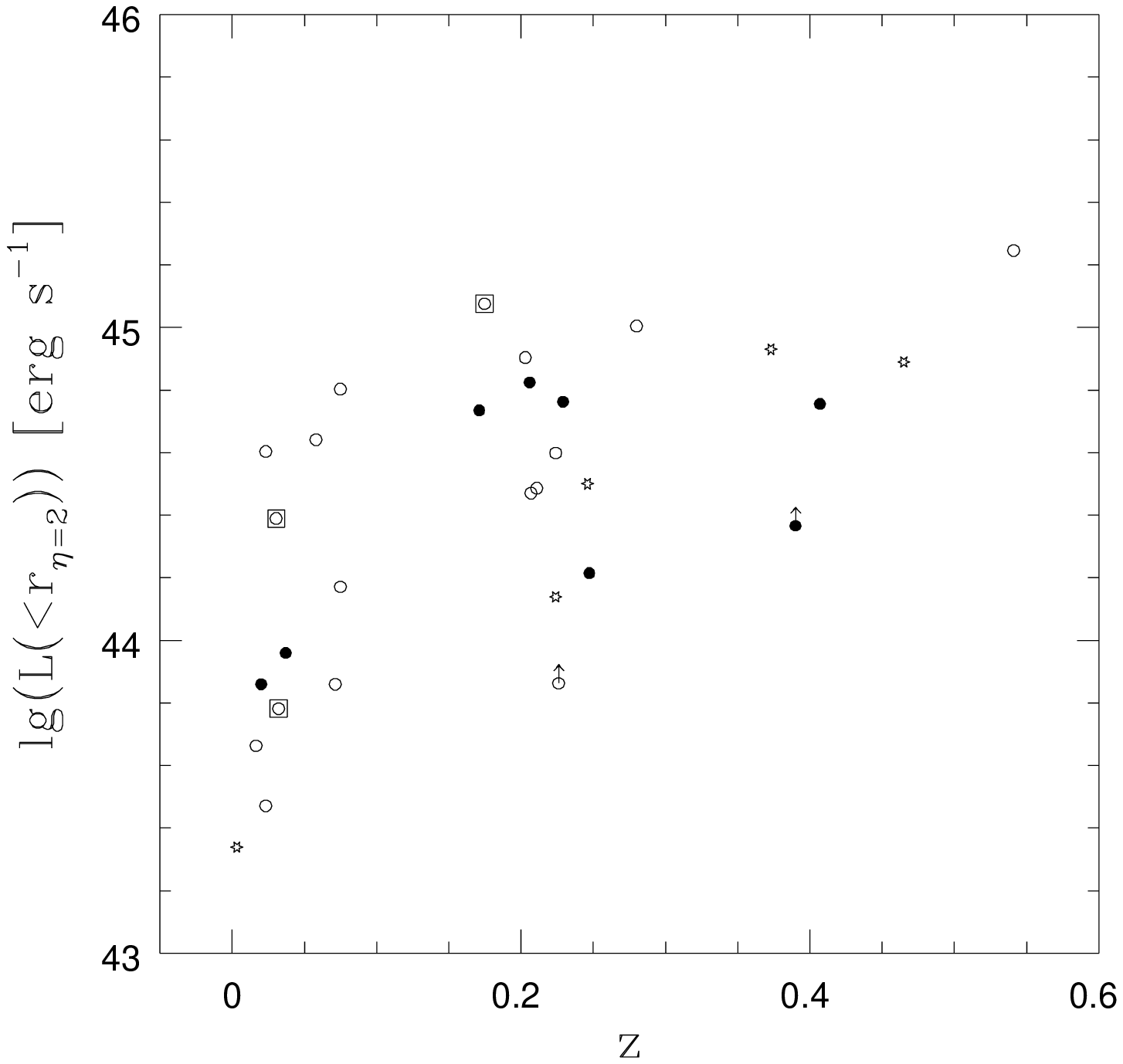}%
\hfill\null}
\caption[h]{X-rax luminosity as a function of $z$. Left panel: apparent
flux of clusters in the HQ sample. The curve is the locus of the clusters having
an x--ray emission 5 time smaller than Abell 1656 (Coma),
assuming a K--correction equal to zero. 
Points are not corrected for absorption or K--correction.
Righ panel: Absolute luminosity for the whole sample.
Absorption and K corrections have been applied
to the data. Literature data are plotted as star points.
Other symbols as Figure 6.}
\end{figure}

\begin{figure}
\hbox{
\epsfysize=8cm
\epsfbox[70 200 470 590]{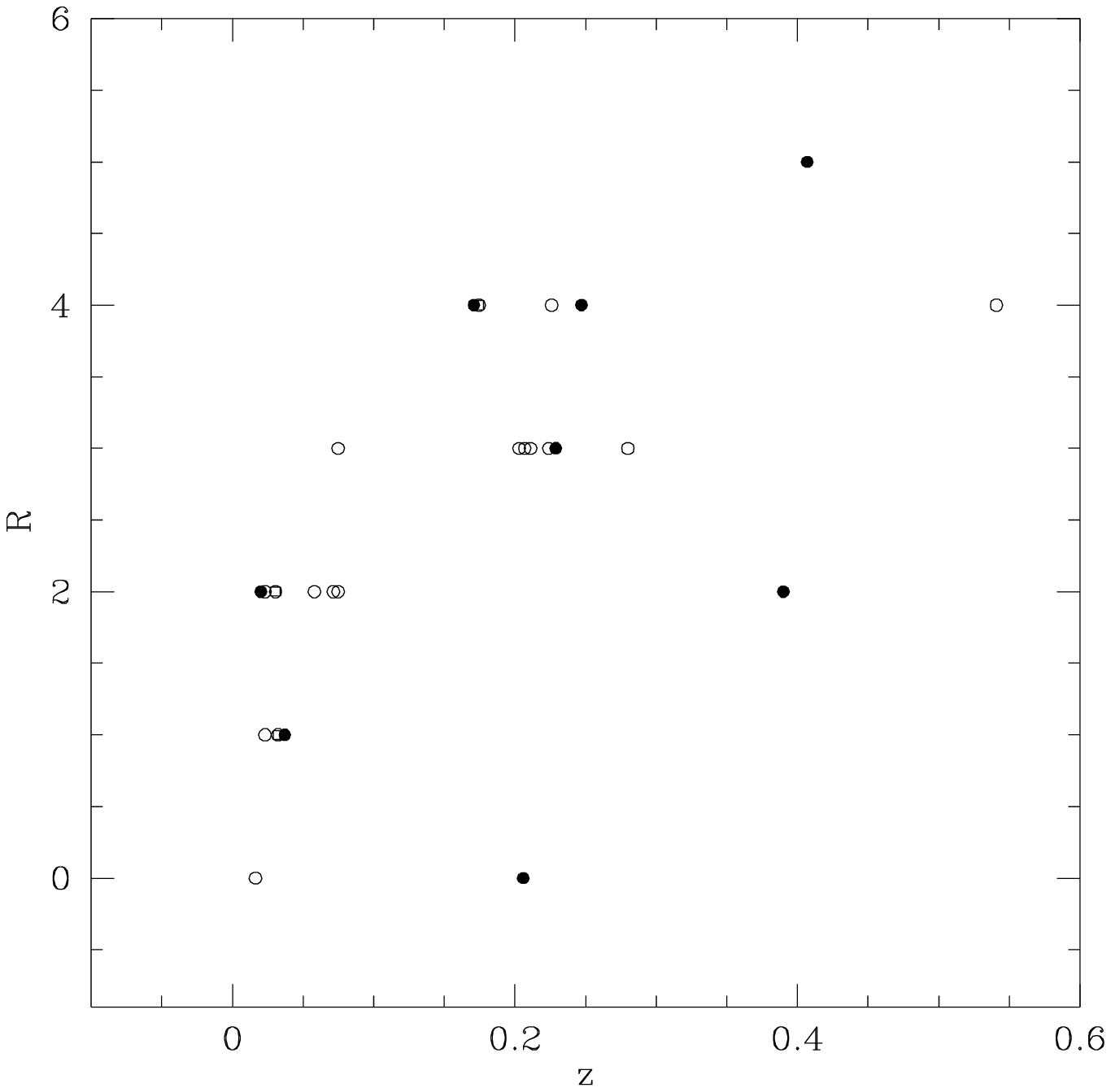}%
\epsfysize=8cm
\epsfbox[50 200 470 590]{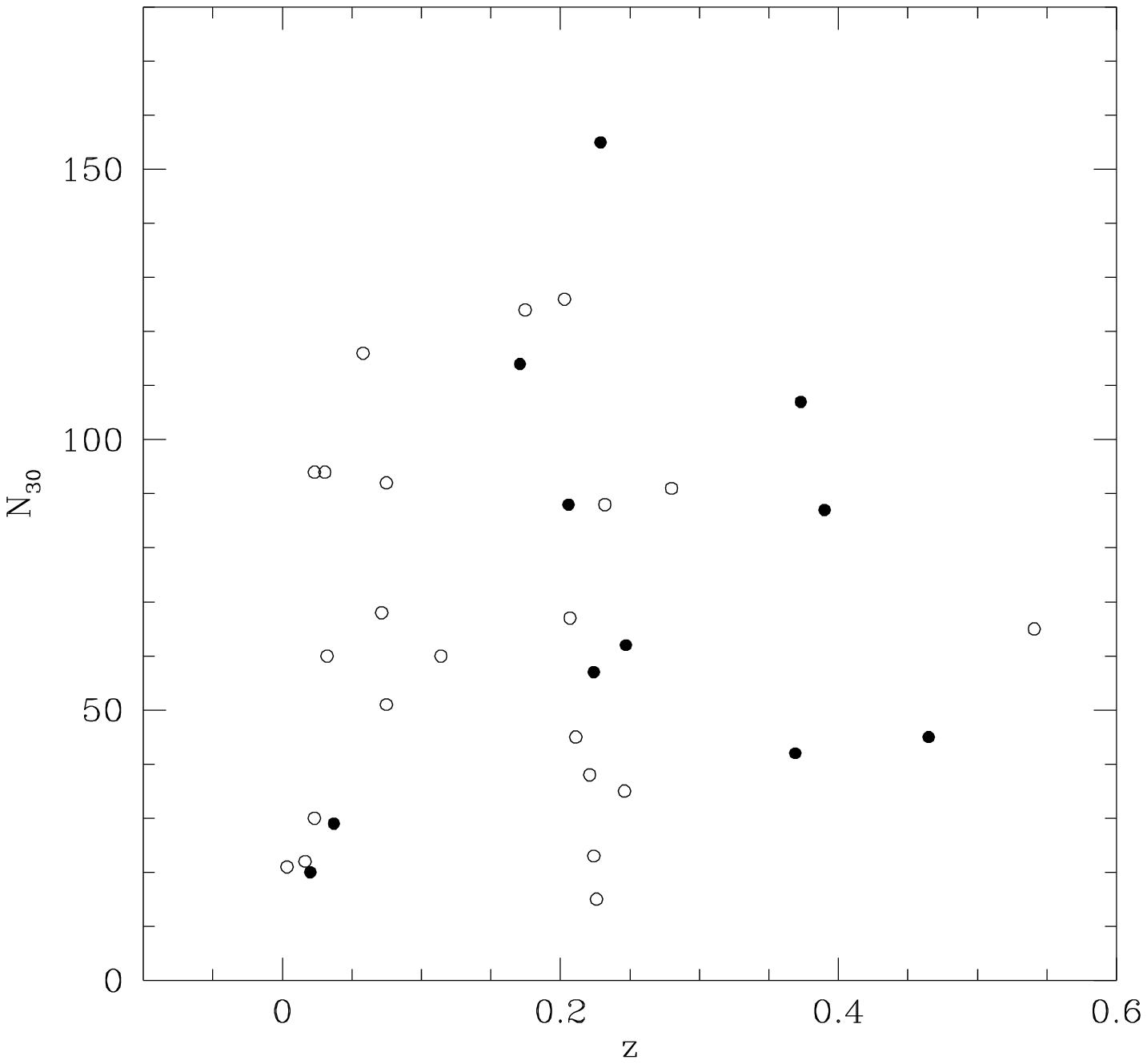}%
\hfill\null}
\caption[h]{Richness as a function of $z$ for the whole sample.
Left panel: ACO richness, right panel:
BO richness. Solid dots are clusters rich in blue galaxies ($f_b>0.1$), 
open dots are clusters poor in blue galaxies.}
\end{figure}

\begin{figure}
\hbox{
\epsfysize=8cm
\epsfbox[40 190 460 590]{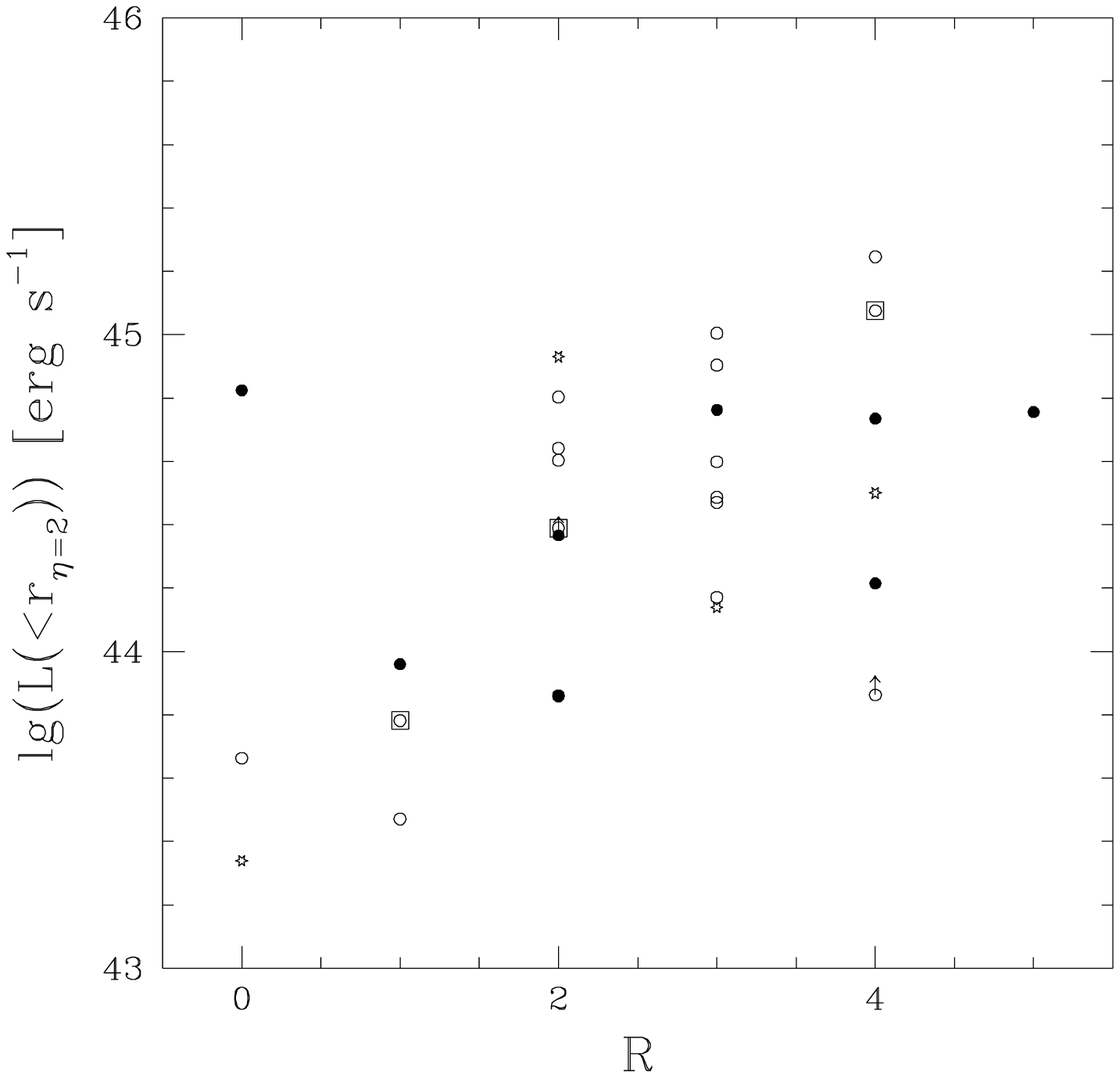}%
\epsfysize=8cm
\epsfbox[40 190 460 590]{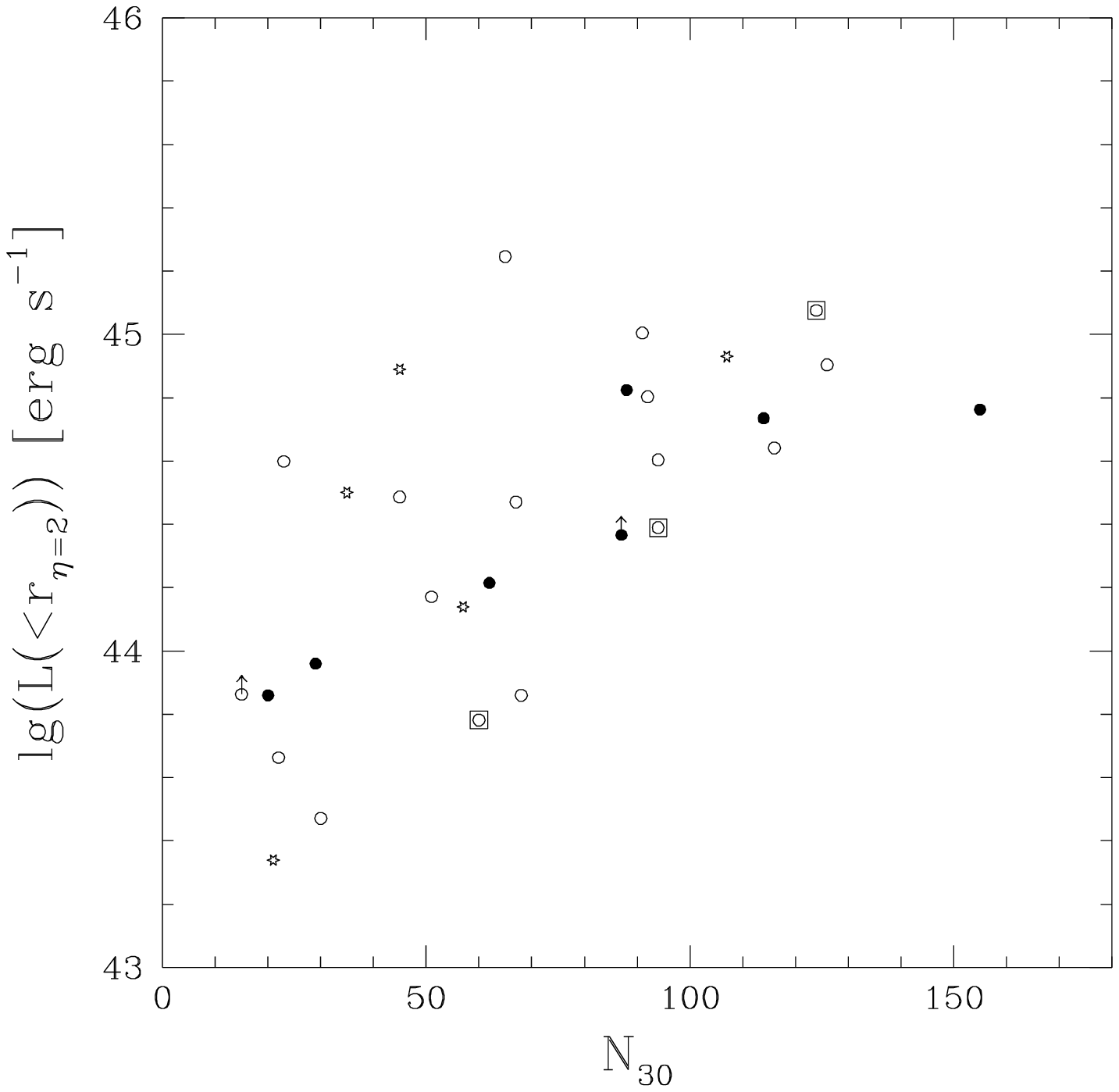}%
\hfill\null}
\caption[h]{X--ray luminosity as a function of the cluster richness
for the whole sample. Left panel:
ACO richness, right panel BO richness. Symbols as Figure 8.}
\end{figure}

\begin{figure}
\hbox{
\epsfysize=8cm
\epsfbox[40 190 460 590]{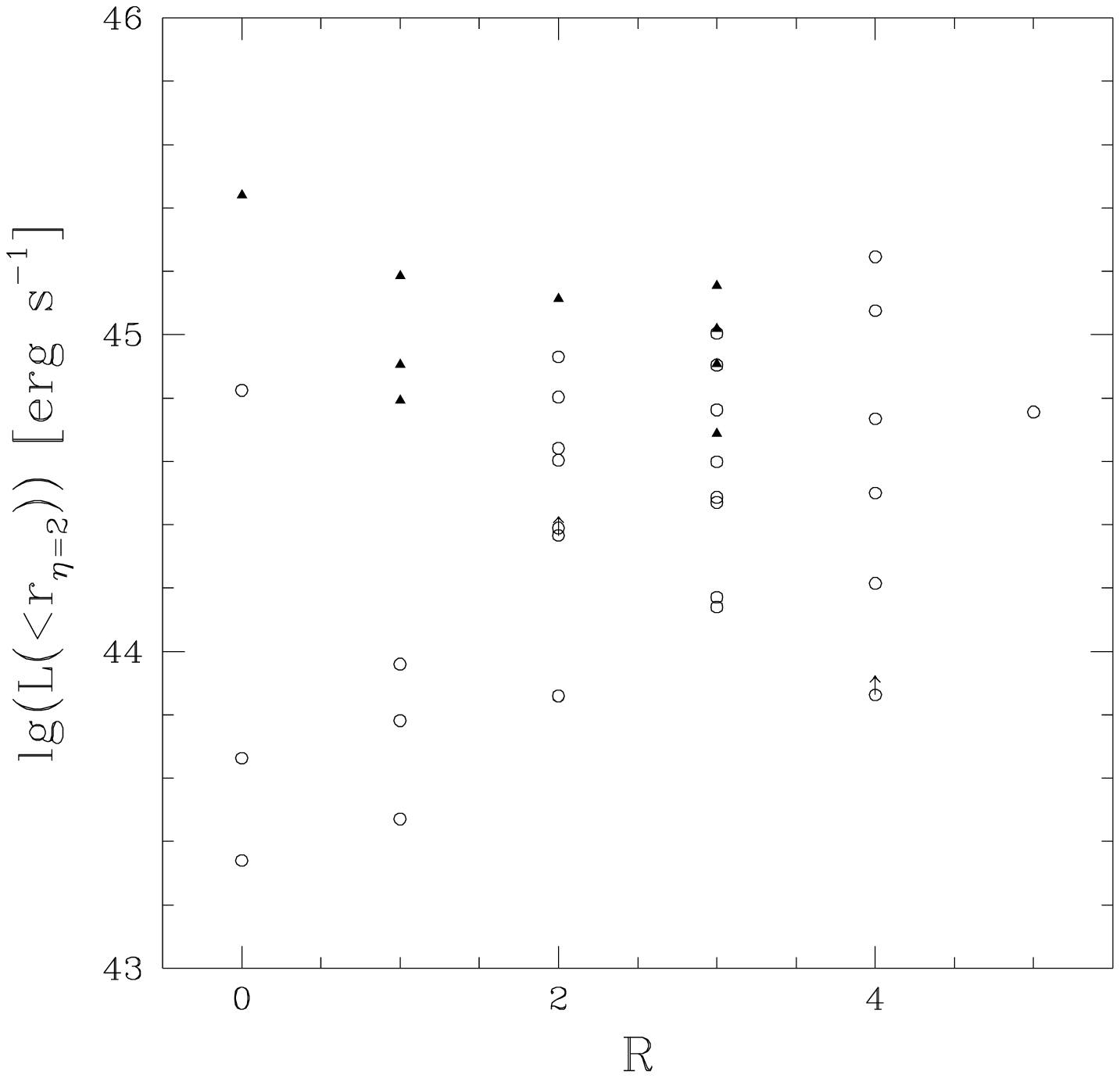}%
\epsfysize=8cm
\epsfbox[40 190 460 590]{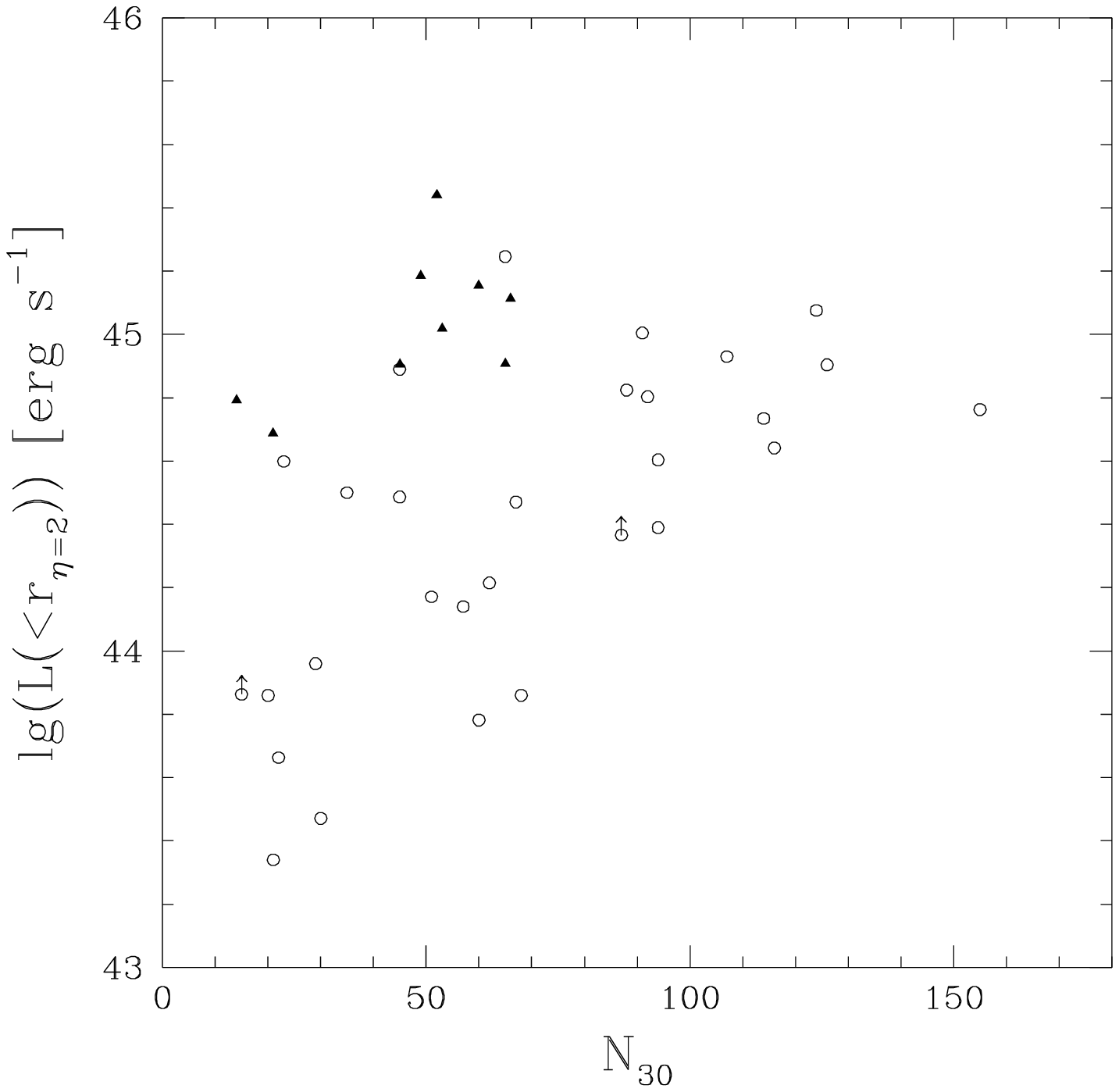}%
\hfill\null}
\caption[h]{X--ray luminosity as a function of the cluster richness,
including the Smail et al. (1998) sample.
Left panel:
ACO richness, right panel BO richness. Open points are the 
optically--selected BO sample, close triangles are the X--ray selected
sample.}
\end{figure}


\begin{references}

\reference{}
Abell G., Corwin H., Olowin R., 1989, ApJS 70, 1

\reference{}
Allen S., Fabian A., 1998, MNRAS 297, L57

\reference{}
Allington-Smith J., Ellis R., Zirbel E., Oemler A., 1993, ApJ 404, 521  

\reference{}
Andreon S., 1994, A\&A 284, 801

\reference{}
Andreon S., 1996, A\&A 314, 763


\reference{}
Andreon S., Davoust E., Heim, A\&A, 1997, 323, 337

\reference{}
Bahcall N., 1974, ApJ 193, 529

\reference{}
Beers T.C., Gebhardt K., Huchra J.P., Forman W., Jones C., Bothun G.D.,
1992, ApJ 400, 410

\reference{}
Biviano A., Katgert P., Mazure A., 1997, A\&A 321, 84

\reference{}
Bothun G., Dressler A., 1986, ApJ 301, 57

\reference{}
Briel U, Henry J., Boeringer H., 1992, A\&A 259, L31

\reference{}
Buote D., Canizares C., 1996, ApJ 457, 565

\reference{}
Butcher H., Oemler A., 1977, ApJ 219, 18

\reference{}
Butcher H., Oemler A., 1984, ApJ 285, 426

\reference{}
Cavaliere A., Fusco-Femiano R., 1976, A\&A 19, 137

\reference{}
Charlot S., Silk J., 1994, ApJ 432, 453

\reference{}
Cirimele G., Nesci R., Trevese D., 1997, ApJ 475, 11

\reference{}
Collins C., Burke D., Romer A., Sharples R., Nicol R., 1997, ApJ 479,
L117

\reference{}
Couch W., Sharples R. 1987, MNRAS 229, 423

\reference{}
David L., Jones C., Forman W., 1996, ApJ 473, 692

\reference{}
Donas J., Milliard B., Laget M., 1995, A\&A, 303, 661

\reference{}
Dressler A., 1980, ApJ 236, 351

\reference{}
Dressler A., Gunn J. 1992, ApJS 78, 1

\reference{}
Dressler A., Oemler A., Couch  W., et al. 1997, ApJ 490, 577 

\reference{}
Edge A. \& Stewart G. 1991, MNRAS 252, 428

\reference{}
Ellis R., 1997, ARA\&A 35, 389

\reference{}
Ellis R., Colless M., Broadhurst T., Heyl J., Glazembrook, K., 1996, MNRAS
280, 235

\reference{}
Ellis R., Smail I., Dressler A., et al. 1997, ApJ 483, 582

\reference{}
Evrard A., 1990, ApJ 363, 349

\reference{}
Evrard A., 1991, MNRAS, 248, 88

\reference{}
Gavazzi G., 1987, ApJ 320, 96


\reference{}
Gunn J., Gott J., 1972, ApJ 176, 1

\reference{}
Jones C., Forman W., 1978, ApJ 224, 1

\reference{}
Jones L., Scharf, C., Ebeling, H., Perlman, E., Wegner, G., Malkan, M.,
\& Horner, D., 1998, ApJ, 495, 100

\reference{}
Henry J, Branduardi G., Fabricant D. et al., 1979, ApJ 234, L15

\reference{}
Henry J., Gioia I., Maccacaro T. et al., 1992, ApJ 386, 408

\reference{}
Huang J.-S., Cowie L., Luppino G., 1998, ApJ 463, 31

\reference{}
Hubble E., Humason M., 1926, ApJ 64, 321

\reference{}
Kauffmann G., 1995, MNRAS 274, 161

\reference{}
Lavery R., Henry J.P., 1988, ApJ 330, 596

\reference{}
Lea S.M., Henry J.P., 1988, ApJ 332, 81

\reference{}
Lilly S., Tresse L., Hammer F., Crampton D., LeFevre O., 1995, ApJ 455, L8

\reference{} 
Maccacaro T., Gioia I, Zamorani G. et al. 1982, Apj 253, 504

\reference{} 
Markevitch M., Vikhlinin A., 1997, ApJ 491, 467 

\reference{}
Mellier Y., Soucail G., Fort B., Mathez G., 1988, A\&A 199, 13

\reference{}
Morrison R., McCammon D., 1983, ApJ, 270,119

\reference{}
Mushotzky R., Scharf C., 1997, ApJ 482, 13

\reference{} 
Koo D., 1981, ApJ 251, L75

\reference{} 
Oemler A., Dressler A, Butcher H., 1997, ApJ 474, 561  

\reference{} 
Petrosian V., 1976, ApJ 209, L1

\reference{} 
Quintana H., Melnick J. 1982, AJ 87, 972

\reference{}
Raymond J.C., Smith B.W., 1977, ApJS, 35, 419

\reference{} 
Rosati P., Della Ceca R., Norman C., Giacconi R., 1998, ApJ 492, L21

\reference{} 
Sadat R., Blanchard A., Guiderdoni B., Silk J., 1998, A\&A 331, L69

\reference{} 
Sandage A., 1988, ARA\&A 26, 561

\reference{} 
Sandage A., Perelmuter J.-M., 1990, ApJ 350, 481

\reference{} 
Sanrom\`a M., Salvador-Sol\'e E., 1990, ApJ 360, 16

\reference{} 
Scaramella R., Zamorani G., Vettolani G., Chincarini G., 1991, AJ 101, 342

\reference{} 
Schindler S., Prieto M., 1997, A\&A 327, 37

\reference{} 
Schindler S., Wambganss J., 1996, A\&A 313, 113

\reference{} 
Smail I., Edge A., Ellis R., Blandford R., 1998, MNRAS, 293, 124

\reference{} 
Soltan A., Henry J., 1983, ApJ 271, 442

\reference{} 
Stanford S., Eisenhardt P., Dickinson M., 1995, ApJ 450, 512

\reference{} 
Stanford S., Eisenhardt P., Dickinson M., 1998, ApJ 492, 461

\reference{} 
Stark A., Gammie C., Wilson R., et al., 1992, ApJS 79, 77

\reference{} 
Vikhlinin A., McNamara B., Forman W., Jones C., Quintana H., 1998, ApJ,
498, L21

\reference{} 
de Vaucouleurs G., 1976

\reference{} 
Wang Q., Ulmer M., 1997, MNRAS, 292, 920 

\reference{}
White D.A., Jones C., Forman W., 1997, MNRAS 292, 419

\reference{} 
Whitmore B., Gilmore D., Jones C., 1993, ApJ 407, 489

\reference{} 
Zwicky F., 1957, Morphological Astronomy, Springer--Verlag

\end{references}
\end{document}